\documentclass[
reprint,
superscriptaddress,  %NOTE: superscriptaddress lists author names separately from institution names.
 amsmath,amssymb,
 aps,
 prd,
]{revtex4-2}

\usepackage{graphicx}% Include figure files
\usepackage{dcolumn}% Align table columns on decimal point
\usepackage{bm}% bold math
\usepackage{amsmath} % For math symbols
\usepackage{hyperref} % For clickable URL links
\usepackage{braket} % For Dirac bra kets
\hypersetup{
    colorlinks=true,
    linkcolor=blue,
    filecolor=magenta,
    citecolor=blue,
    urlcolor=blue
    }
\usepackage[all]{hypcap} % For linking hyprefs to top of figure
\usepackage{graphicx} % For figures
\usepackage{gensymb} % Note for degree symbol
\usepackage{csvsimple}
\usepackage{siunitx}
\usepackage{array}
\usepackage{booktabs}
\usepackage{multirow}
\usepackage[table]{xcolor}
\usepackage{lineno}
\usepackage{placeins} % For \FloatBarrier

\usepackage[export]{adjustbox}

\begin{document}

\preprint{APS/123-QED}

\title{Longitudinal Spin Transfer to \texorpdfstring{$\Lambda$}{Lambda} Hyperons in Semi-Inclusive Deep Inelastic Scattering with CLAS12}

\newcommand*{\ANL}{Argonne National Laboratory, Argonne, Illinois 60439}
\newcommand*{\ANLindex}{1}
\affiliation{\ANL}
\newcommand*{\CANISIUS}{Canisius University, Buffalo, NY}
\newcommand*{\CANISIUSindex}{2}
\affiliation{\CANISIUS}
\newcommand*{\SACLAY}{IRFU, CEA, Universit\'{e} Paris-Saclay, F-91191 Gif-sur-Yvette, France}
\newcommand*{\SACLAYindex}{3}
\affiliation{\SACLAY}
\newcommand*{\CNU}{Christopher Newport University, Newport News, Virginia 23606}
\newcommand*{\CNUindex}{4}
\affiliation{\CNU}
\newcommand*{\UCONN}{University of Connecticut, Storrs, Connecticut 06269}
\newcommand*{\UCONNindex}{5}
\affiliation{\UCONN}
\newcommand*{\DUKE}{Duke University, Durham, North Carolina 27708-0305}
\newcommand*{\DUKEindex}{6}
\affiliation{\DUKE}
\newcommand*{\DUQUESNE}{Duquesne University, 600 Forbes Avenue, Pittsburgh, PA 15282 }
\newcommand*{\DUQUESNEindex}{7}
\affiliation{\DUQUESNE}
\newcommand*{\FU}{Fairfield University, Fairfield CT 06824}
\newcommand*{\FUindex}{8}
\affiliation{\FU}
\newcommand*{\FERRARAU}{Universita' di Ferrara , 44121 Ferrara, Italy}
\newcommand*{\FERRARAUindex}{9}
\affiliation{\FERRARAU}
\newcommand*{\FIU}{Florida International University, Miami, Florida 33199}
\newcommand*{\FIUindex}{10}
\affiliation{\FIU}
\newcommand*{\GWUI}{The George Washington University, Washington, DC 20052}
\newcommand*{\GWUIindex}{11}
\affiliation{\GWUI}
\newcommand*{\GSIFFN}{GSI Helmholtzzentrum fur Schwerionenforschung GmbH, D-64291 Darmstadt, Germany}
\newcommand*{\GSIFFNindex}{12}
\affiliation{\GSIFFN}
\newcommand*{\INFNFE}{INFN, Sezione di Ferrara, 44100 Ferrara, Italy}
\newcommand*{\INFNFEindex}{13}
\affiliation{\INFNFE}
\newcommand*{\INFNFR}{INFN, Laboratori Nazionali di Frascati, 00044 Frascati, Italy}
\newcommand*{\INFNFRindex}{14}
\affiliation{\INFNFR}
\newcommand*{\INFNGE}{INFN, Sezione di Genova, 16146 Genova, Italy}
\newcommand*{\INFNGEindex}{15}
\affiliation{\INFNGE}
\newcommand*{\INFNRO}{INFN, Sezione di Roma Tor Vergata, 00133 Rome, Italy}
\newcommand*{\INFNROindex}{16}
\affiliation{\INFNRO}
\newcommand*{\INFNTUR}{INFN, Sezione di Torino, 10125 Torino, Italy}
\newcommand*{\INFNTURindex}{17}
\affiliation{\INFNTUR}
\newcommand*{\INFNCAT}{INFN, Sezione di Catania, 95123 Catania, Italy}
\newcommand*{\INFNCATindex}{18}
\affiliation{\INFNCAT}
\newcommand*{\INFNPAV}{INFN, Sezione di Pavia, 27100 Pavia, Italy}
\newcommand*{\INFNPAVindex}{19}
\affiliation{\INFNPAV}
\newcommand*{\ORSAY}{Universit\'{e} Paris-Saclay, CNRS/IN2P3, IJCLab, 91405 Orsay, France}
\newcommand*{\ORSAYindex}{20}
\affiliation{\ORSAY}
\newcommand*{\JMU}{James Madison University, Harrisonburg, Virginia 22807}
\newcommand*{\JMUindex}{21}
\affiliation{\JMU}
\newcommand*{\KNU}{Kyungpook National University, Daegu 41566, Republic of Korea}
\newcommand*{\KNUindex}{22}
\affiliation{\KNU}
\newcommand*{\LAMAR}{Lamar University, 4400 MLK Blvd, PO Box 10046, Beaumont, Texas 77710}
\newcommand*{\LAMARindex}{23}
\affiliation{\LAMAR}
\newcommand*{\MIT}{Massachusetts Institute of Technology, Cambridge, Massachusetts  02139-4307}
\newcommand*{\MITindex}{24}
\affiliation{\MIT}
\newcommand*{\MISS}{Mississippi State University, Mississippi State, MS 39762-5167}
\newcommand*{\MISSindex}{25}
\affiliation{\MISS}
\newcommand*{\UNH}{University of New Hampshire, Durham, New Hampshire 03824-3568}
\newcommand*{\UNHindex}{26}
\affiliation{\UNH}
\newcommand*{\NMSU}{New Mexico State University, PO Box 30001, Las Cruces, NM 88003, USA}
\newcommand*{\NMSUindex}{27}
\affiliation{\NMSU}
\newcommand*{\NSU}{Norfolk State University, Norfolk, Virginia 23504}
\newcommand*{\NSUindex}{28}
\affiliation{\NSU}
\newcommand*{\OHIOU}{Ohio University, Athens, Ohio  45701}
\newcommand*{\OHIOUindex}{29}
\affiliation{\OHIOU}
\newcommand*{\ODU}{Old Dominion University, Norfolk, Virginia 23529}
\newcommand*{\ODUindex}{30}
\affiliation{\ODU}
\newcommand*{\JLUGiessen}{II Physikalisches Institut der Universitaet Giessen, 35392 Giessen, Germany}
\newcommand*{\JLUGiessenindex}{31}
\affiliation{\JLUGiessen}
\newcommand*{\ROMAII}{Universita' di Roma Tor Vergata, 00133 Rome Italy}
\newcommand*{\ROMAIIindex}{32}
\affiliation{\ROMAII}
\newcommand*{\MSU}{Skobeltsyn Institute of Nuclear Physics, Lomonosov Moscow State University, 119234 Moscow, Russia}
\newcommand*{\MSUindex}{33}
\affiliation{\MSU}
\newcommand*{\SCAROLINA}{University of South Carolina, Columbia, South Carolina 29208}
\newcommand*{\SCAROLINAindex}{34}
\affiliation{\SCAROLINA}
\newcommand*{\TEMPLE}{Temple University,  Philadelphia, PA 19122 }
\newcommand*{\TEMPLEindex}{35}
\affiliation{\TEMPLE}
\newcommand*{\JLAB}{Thomas Jefferson National Accelerator Facility, Newport News, Virginia 23606}
\newcommand*{\JLABindex}{36}
\affiliation{\JLAB}
\newcommand*{\ULS}{Universidad de La Serena}
\newcommand*{\ULSindex}{37}
\affiliation{\ULS}
\newcommand*{\UTFSM}{Universidad T\'{e}cnica Federico Santa Mar\'{i}a, Casilla 110-V Valpara\'{i}so, Chile}
\newcommand*{\UTFSMindex}{38}
\affiliation{\UTFSM}
\newcommand*{\BRESCIA}{Universit`{a} degli Studi di Brescia, 25123 Brescia, Italy}
\newcommand*{\BRESCIAindex}{39}
\affiliation{\BRESCIA}
\newcommand*{\UCR}{University of California Riverside, 900 University Avenue, Riverside, CA 92521, USA}
\newcommand*{\UCRindex}{40}
\affiliation{\UCR}
\newcommand*{\GLASGOW}{University of Glasgow, Glasgow G12 8QQ, United Kingdom}
\newcommand*{\GLASGOWindex}{41}
\affiliation{\GLASGOW}
\newcommand*{\YORK}{University of York, York YO10 5DD, United Kingdom}
\newcommand*{\YORKindex}{42}
\affiliation{\YORK}
\newcommand*{\VIRGINIA}{University of Virginia, Charlottesville, Virginia 22901}
\newcommand*{\VIRGINIAindex}{43}
\affiliation{\VIRGINIA}
\newcommand*{\WM}{College of William and Mary, Williamsburg, Virginia 23187-8795}
\newcommand*{\WMindex}{44}
\affiliation{\WM}
\newcommand*{\YEREVAN}{Yerevan Physics Institute, 375036 Yerevan, Armenia}
\newcommand*{\YEREVANindex}{45}
\affiliation{\YEREVAN}

\newcommand*{\NOWJLAB}{Thomas Jefferson National Accelerator Facility, Newport News, Virginia 23606}
\newcommand*{\NOWDOTA}{DOTA, ONERA, Universit\'{e} Paris-Saclay, 91120, Palaiseau, France}
\newcommand*{\NOWGWUI}{The George Washington University, Washington, DC 20052}

\author {M.~McEneaney} 
\email{matthew.mceneaney@duke.edu}
\affiliation{\DUKE}
\author {A.~Vossen}
\email{anselm.vossen@duke.edu}
\affiliation{\DUKE}
\affiliation{\JLAB}
\author {A.~Acar} 
\affiliation{\YORK}
\author {P.~Achenbach} 
\affiliation{\JLAB}
\author {J.S.~Alvarado} 
\affiliation{\ORSAY}
\author {M.~Amaryan} 
\affiliation{\ODU}
\author {W.R.~Armstrong} 
\affiliation{\ANL}
\author {H.~Atac} 
\affiliation{\TEMPLE}
\author {N.A.~Baltzell} 
\affiliation{\JLAB}
\author {L.~Barion} 
\affiliation{\INFNFE}
% ...

\collaboration{The CLAS Collaboration}\noaffiliation

\date{\today}

\begin{abstract}
    The polarization of $\Lambda$ hyperons is preserved in the angular distribution of their decay products.  This property allows one to study the spin structure of the $\Lambda$.  In Semi-Inclusive Deep Inelastic Scattering where a high energy lepton interacts with a nucleon target and one or more hadrons and the scattered lepton are detected in the final state, the probability for a struck quark to impart the polarization of the lepton to the $\Lambda$ may be measured.  In particular, in electron-proton scattering this quantity may be related to the longitudinal light quark polarization of the $\Lambda$.  Currently, limited experimental data cannot discriminate between different models of $\Lambda$ spin structure.  This work reports on the measurement of the longitudinal spin transfer $D^{\Lambda}_{LL'}$ to the $\Lambda$ using data taken by the CLAS12 spectrometer at Jefferson Lab with a $10.6$~GeV longitudinally polarized electron beam and an unpolarized hydrogen target.  This measurement is the most precise to date, and, in comparison with theory predictions, it offers valuable insight into the relative dominance of current and target fragmentation in $\Lambda$ production.
\end{abstract}

%\keywords{Suggested keywords}%Use showkeys class option if keyword
                              %display desired
\maketitle

%\tableofcontents

% Introduction
\section{Introduction} \label{sec:Introduction}
Understanding the internal dynamics of the nucleon and other hadrons remains one of the most complex unsolved challenges in modern physics.  The quarks within a hadron are held together by the strong force, one of the four fundamental forces.  Quantum Chromodynamics (QCD) gives us a theoretical framework that has quite successfully described many physical observations~\cite{Halzen_and_Martin}.  However, hadrons are strongly coupled systems and, hence, are not perturbatively calculable.  Fortunately, the asymptotic freedom of QCD at short distances allows one to analyze the hadron dynamics with a high energy probe such as an electron since its de Broglie wavelength $\lambda = h / p$ is inversely related to its energy~\cite{Halzen_and_Martin}.

Processes such as Semi-Inclusive Deep Inelastic Scattering (SIDIS)
\begin{equation}
    \ell(l) + N(P) \rightarrow \ell'(l') + h(P_h) + X,
    \label{eq:introduction_sidis}
\end{equation}
where a high energy lepton $\ell$ collides with a target nucleon $N$ and one or more hadrons $h$ are detected in the final state are ideally suited to studying non-perturbative quantities in QCD.  $l$, $l'$, $P$, and $P_h$, respectively, denote the 4-momenta of the incoming and outgoing lepton, the target nucleon, and a final state hadron.  The 4-momentum of the virtual photon $\gamma^*$ exchanged between the incoming lepton and the target nucleon is denoted $q = l - l'$.  Some kinematic variables used to characterize the interactions are defined as follows
%----- Eq: Kinematics -----%
\begin{subequations}
	\label{eq:introduction_kinematics}
	\begin{align}
        \label{eq:introduction_kinematics:q2}
		Q^2 &= -q^{\mu}q_{\mu},\\
        \label{eq:introduction_kinematics:w}
		W^2 &= (P^{\mu}+q^{\mu})^2,\\
        \label{eq:introduction_kinematics:x}
		x &= \frac{Q^2}{2 P^{\mu}q_{\mu}},\\
        \label{eq:introduction_kinematics:y}
		y &= \frac{P^{\mu}q_{\mu}}{P^{\mu}l_{\mu}},\\
        \label{eq:introduction_kinematics:z}
		z_h &= \frac{P^{\mu}P_{h, \mu}}{P^{\mu}q_{\mu}},\\
        \label{eq:introduction_kinematics:xf}
		x_{F  h} &= \frac{2 \vec{P}_h \cdot \vec{q}}{W \lvert q \rvert},\\
        \label{eq:introduction_kinematics:phi_h}
		\cos{\phi_h} &= \frac{(\vec{q} \times \vec{P}_h) \cdot (\vec{q} \times \vec{l'})}{|\vec{q} \times \vec{P}_h| |\vec{q} \times \vec{l'}|}.
	\end{align}
\end{subequations}
$Q^2$ is the negative 4-momentum squared of the $\gamma^*$ momentum and describes the virtuality of the interaction. $W$ is the final state energy in the $\gamma^* N$ center-of-mass (CM) frame, $x$ is the longitudinal light-cone momentum fraction of the target nucleon carried by the struck quark, and $y$ is the fraction of the incoming electron energy carried by the virtual photon.  The remaining variables are specific to the hadron $h$ of interest detected in the final state.  $z_h$ describes the energy fraction of the hadron relative to the struck quark and $x_{F h}$ is the longitudinal momentum fraction of the hadron relative to the maximum value $W/2$ allowed by momentum conservation in the $\gamma^*N$ CM frame.  Finally, $\phi_h$ is the azimuthal angle of the hadron momentum about the virtual photon momentum in the $\gamma^*N$ CM frame.  Importantly, the $x_{F h}$ variable is used to distinguish between hadrons produced from the struck quark, the Current Fragmentation Region (CFR), and from the remaining quarks within the target nucleon, the Target Fragmentation Region (TFR).  Positive $x_{F h}$ values are associated with hadrons traveling in the forward-going virtual photon direction and negative values are associated with hadrons traveling in the backward-going target nucleon direction in the $\gamma^*N$ CM frame.

\begin{figure}[ht!]
\centering
    \includegraphics[width=0.35\textwidth]{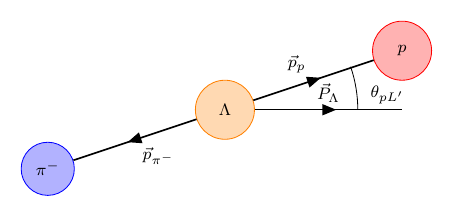}
\caption{The $\Lambda \rightarrow p\pi^{-}$ decay in the $\Lambda$ rest frame is shown with the $\Lambda$ spin axis $\vec{P}_{\Lambda}$ retained from the $\gamma^*N$ CM frame.}
\label{fig:lambda_decay}
\end{figure}

While unpolarized SIDIS gives access to unpolarized observables in QCD, there still remain transverse momentum and polarization dependent aspects of hadron structure that can only be accessed through polarized SIDIS measurements.  One must consider these aspects of QCD as well in order to construct a complete picture of a hadronic QCD system.  In SIDIS, observables dependent on transverse momentum and polarization often manifest in asymmetries between the cross sections produced from opposite initial lepton probe helicity and target nucleon spin configurations.  The $\Lambda$ hyperon lends itself naturally to these studies by virtue of its self-analyzing weak decay $\Lambda \rightarrow p\pi^{-}$ shown in Fig.~\ref{fig:lambda_decay}.  In fact, the large spontaneous transverse polarization of $\Lambda$ hyperons produced in $p+Be$ collisions~\cite{PhysRevLett.36.1113} over $45$ years ago was the first indication that transverse spin effects played a significant role in hadron physics.  The polarization of the $\Lambda$ is preserved in the cross section of the decay protons
\begin{equation}
     \frac{dN}{d\Omega_p} \propto 1 + \alpha_{\Lambda}\vec{\text{P}}_{\Lambda} \cdot \hat{n}_p.
    \label{eq:introduction_lambda_decay_xs}
\end{equation}
$\alpha_{\Lambda}=0.747\pm0.009$ is the asymmetry parameter for the weak decay of the $\Lambda$ in its CM frame~\cite{ParticleDataGroup:2024cfk} and $\vec{\text{P}}_{\Lambda} \cdot \hat{n}_p$ is the polarization of the $\Lambda$ along the proton momentum in the $\Lambda$ rest frame~\cite{Airapetian_2006,Guan_2019}.

In this paper, we set out to expand upon our previous work~\cite{McEneaney_2022} and to make a clean and precise measurement of the $\Lambda$ longitudinal spin transfer coefficient $D^{\Lambda}_{LL'}$.  This quantity describes the probability for a struck quark to impart its longitudinal spin to the $\Lambda$ in SIDIS, and, as discussed below, for our selected data, this may be assumed to come primarily from scattering off light ($u$, $d$) quarks within the target nucleon.  Measurement of this quantity also offers a clear test of the dominant $\Lambda$ production mechanism in the CLAS12 kinematic phase space, which provides significant insight for future studies.

% From Eq.~\ref{eq:introduction_lambda_decay_xs} we see that the $\Lambda$ decay cross section is itself dependent on the $\Lambda$ polarization $\text{P}_{\Lambda}$, which we may relate to the longitudinally polarized $\Lambda$ production cross section with a polarized electron beam~\cite{PhysRevD.95.074026}:
% \begin{align}
%     \text{P}_{\Lambda}& = \frac{N^{\Lambda+}-N^{\Lambda-}}{N^{\Lambda+}+N^{\Lambda-}} \propto \frac{d\sigma_{LUL}}{d\sigma_{UUU}}.
%     \label{eq:introduction_lambda_polarization_asymmetry}
% \end{align}
% We can further express the $\Lambda$ polarization $\text{P}_{\Lambda}$ in terms of Parton Distribution Functions (PDFs) and Fragmentation Functions (FFs):
% \begin{align}
%     \text{P}_{\Lambda} = \lambda_{\ell} \hspace{3pt} \frac{y\big{(}1 - \frac{1}{2}y \big{)}}{1 - y + \frac{1}{2}y^2} \hspace{3pt} \frac{\sum_a e^2_a f^a_1(x) G^a_1(z)}{\sum_a e^2_a f^a_1(x) D^a_1(z)},
%     \label{eq:introduction_lambda_polarization_factorization}
% \end{align}
% where the sum over $a$ is a sum over quark and antiquark flavors and $f_1$ is the unpolarized PDF and $D_1$ and $G_1$ are the unpolarized and helicity FFs respectively, following the notational conventions of Ref.~\cite{Mulders_1996}.

\subsection{The Polarized \texorpdfstring{$\Lambda$}{Lambda} Production Cross Section}\label{sec:introduction:the_polarized_lambda_production_XS}

As shown in Ref.~\cite{PhysRevD.95.074026}, the only term of the $\Lambda$ lepto-production differential cross section with an unpolarized target that is dependent on both the longitudinal helicity state of the incoming lepton beam $\lambda_{\ell}$ and of the $\Lambda$ longitudinal spin state $S_{\Lambda_{||}}$ is
\begin{multline}
    d\sigma_{LUL} \propto S_{\Lambda_{||}} \lambda_{\ell} \bigg{[} y\bigg{(} 1 - \frac{1}{2} y\bigg{)} F_{LUL} \\
    + y\sqrt{1-y}\cos{\phi_{\Lambda}}F^{\cos{\phi_{\Lambda}}}_{LUL} \bigg{]}.
    \label{eq:introduction_lambda_xs}
\end{multline}
Here, $\phi_{\Lambda}$ is the azimuthal angle of the $\Lambda$ momentum about the $\gamma^*$ momentum in the $\gamma^{*}N$ CM frame defined by Eq.~\ref{eq:introduction_kinematics:phi_h} and $F^{Mod}_{ABC}=F^{Mod}_{ABC}(x,z,P^2_{h\perp})$ are structure functions of different angular modulations denoted by the superscripts.  The subscripts correspond to beam, target, and $\Lambda$ polarization states, respectively.  As shown in Ref.~\cite{PhysRevD.95.074026}, integrating over the transverse momentum $P_{h\perp}$ of the $\Lambda$, leaves us with just the first structure function $F_{LUL}$, which may be expressed in terms of Parton Distribution Functions (PDFs) and Fragmentation Functions (FFs) as
\begin{equation}
    F_{LUL}(x,z) = x \sum_a e^2_a f^a_1(x) G^{\Lambda,a}_1(z),
    \label{eq:introduction_lambda_xs_factorized}
\end{equation}
where the sum runs over the quark flavors $a$ and $e_a$ is the the charge of the quark of flavor $a$.  $f_1$ is the unpolarized PDF and $G_1$ is the helicity FF, following the notational conventions of Ref.~\cite{Mulders_1996}.  The only unpolarized differential cross section term that survives the $P_{h\perp}$ integration is given by~\cite{PhysRevD.57.5780}
\begin{align}
\begin{split}
    d\sigma_{UUU}& \propto \bigg{(} 1 - y + \frac{1}{2}y^2 \bigg{)} F_{UUU} \\
    & \propto \bigg{(} 1 - y + \frac{1}{2}y^2 \bigg{)} x \sum_a e^2_a f^a_1(x) D^{\Lambda,a}_1(z),
    \label{eq:introduction_lambda_xs_unpol}
\end{split}
\end{align}
and contains the unpolarized FF $D_1$.

The most basic and extensively studied FF is the unpolarized FF $D_1^{h,a}(z)$ describing the probability for an unpolarized quark of flavor $a$ to fragment into an unpolarized hadron $h$~\cite{METZ2016136}.  However, one may further refine PDFs and FFs by introducing helicity state dependence.  For example, we may define the probability for a quark with flavor $a$ of positive helicity $+$ to fragment into a hadron $h$ of the same helicity as $D_1^{h+,a+}$ and to fragment into a hadron of opposite helicity $-$ as $D_1^{h-,a+}$.  Assuming these probabilities are symmetric for a quark of negative helicity fragmenting into a hadron of the same or opposite helicity, $D_1^{h-,a-}=D_1^{h+,a+}$ and $D_1^{h+,a-}=D_1^{h-,a+}$, the unpolarized FF may then be expressed as $D_1^{h,a}=D_1^{h+,a+}+D_1^{h-,a+}$ and we may define the helicity-dependent FF $G_1^{h,a}=D_1^{h+,a+}-D_1^{h-,a+}$~\cite{Mulders_1996,Airapetian_2006}.  $G_1^{h,a}$ describes the probability to produce a longitudinally polarized hadron from a longitudinally polarized quark.

Recall from Eq.~\ref{eq:introduction_lambda_decay_xs} that the $\Lambda$ decay cross section is itself dependent on the $\Lambda$ polarization $\text{P}_{\Lambda}$ which, fortuitously, is an asymmetry we can relate to the $P_{h\perp}$-integrated cross sections from Eqs.~\ref{eq:introduction_lambda_xs} and~\ref{eq:introduction_lambda_xs_unpol} as
\begin{align}
    \text{P}_{\Lambda}& = \frac{N^{\Lambda+}-N^{\Lambda-}}{N^{\Lambda+}+N^{\Lambda-}} \propto \frac{d\sigma_{LUL}}{d\sigma_{UUU}}.
    \label{eq:introduction_lambda_polarization_asymmetry}
\end{align}
Thus, we can express the $\Lambda$ polarization $\text{P}_{\Lambda}$ in terms of PDFs and FFs
\begin{align}
    \text{P}_{\Lambda} = \lambda_{\ell} \hspace{3pt} \frac{y\big{(}1 - \frac{1}{2}y \big{)}}{1 - y + \frac{1}{2}y^2} \hspace{3pt} \frac{\sum_a e^2_a f^a_1(x) G^{\Lambda,a}_1(z)}{\sum_a e^2_a f^a_1(x) D^{\Lambda,a}_1(z)}.
    \label{eq:introduction_lambda_polarization_factorization}
\end{align}

\subsection{The Longitudinal Spin Transfer Coefficient \texorpdfstring{$D^{\Lambda}_{LL'}$}{DLL}}\label{sec:introduction:the_longitudinal_spin_transfer_coefficient}

In SIDIS, the incoming electron interacts with a valence quark in the target nucleon by means of a virtual photon.  Since the photon has spin 1, a longitudinally polarized beam will preferentially select a quark of the opposite polarization.  This means that the outgoing quark will have the same helicity as the virtual photon and will have some chance of imparting its polarization to a hadron in fragmentation.  With this more intuitive formulation, we can write the $\Lambda$ polarization from Eq.~\ref{eq:introduction_lambda_polarization_factorization} as
\begin{equation}
    \text{P}_{\Lambda} = \lambda_{\ell} D(y) D_{LL'}^{\Lambda},
    \label{eq:lambda_polarization}
\end{equation}
where $\lambda_{\ell}$ is the helicity of the incident electron and $D(y)=\frac{y(1-\frac{1}{2}y)}{1-y+\frac{1}{2}y^2}$ is a depolarization factor that takes into account the loss of polarization from the incident electron to the virtual photon.  The final ratio of PDFs and FFs in Eq.~\ref{eq:introduction_lambda_polarization_factorization} we denote by $D_{LL'}^{\Lambda}$.  This is the longitudinal spin transfer coefficient, which encodes the probability for the struck quark in the target nucleon to impart its polarization to the produced $\Lambda$.  This means that the angular distribution of protons coming from the $\Lambda$ decay is
\begin{equation}
    \frac{dN}{d\Omega_p} \propto 1 + \alpha_{\Lambda} \lambda_{\ell} D(y) D_{LL'}^{\Lambda} \cos{\theta_{pL'}},
    \label{eq:lambda_xs_sidis}
\end{equation}
where the angle $\cos{\theta_{pL'}}$ is taken between the proton momentum and $\Lambda$ spin quantization axis $L'$ in the $\Lambda$ rest frame.  In principle, one may analyze the $\Lambda$ polarization along any choice of the axis $L'$.  Some authors differ in their definition of the longitudinal axis~\cite{Mulders_1996,PhysRevD.61.014007,PhysRevD.65.034004}.  Hence, for our analysis we present results for two different choices of the longitudinal polarization axis: along the virtual photon $\gamma^{*}$ momentum and along the $\Lambda$ momentum direction retained from the $\gamma^*N$ CM frame.
%%%%% FIG: LAMBDA DECAY %%%%%

We now motivate the interpretation of $D^{\Lambda}_{LL'}$ as the probability of spin transfer from the struck quark to the produced $\Lambda$.
% From Eq.~\ref{eq:introduction_lambda_polarization_factorization},
% \begin{equation}
%     D^{\Lambda}_{LL'}(x,z) = \frac{\sum_a e^2_a f^a_1(x) G^{\Lambda,a}_1(z)}{\sum_a e^2_a f^a_1(x) D^{\Lambda,a}_1(z)},
%     \label{eq:introduction_quark_content}
% \end{equation}
Since our CLAS12 data do not span a wide range of $Q^2$ and exhibit almost no $Q^2$ dependence, we have assumed that the PDFs and FFs vary slowly with $Q^2$.  Integrating the spin transfer coefficient as formulated in Eq.~\ref{eq:introduction_lambda_polarization_factorization} over all possible momentum fractions $x$ of the struck quark and inserting an identity, we obtain~\cite{Airapetian_2006}
\begin{equation}
    D^{\Lambda}_{LL'}(z) \simeq \sum_{a} \frac{G^{\Lambda,a}_{1}(z)}{D^{\Lambda,a}_{1}(z)} p^{\Lambda,a},
    \label{eq:x_integration}
\end{equation}
where the purity $p^{\Lambda,a}$ is the probability that a $\Lambda$ was produced at a fractional energy $z$ from a quark of flavor $a$
\begin{equation}
    p^{\Lambda,a} = \int \frac{e^2_a f^a_1(x ) D^{\Lambda,a}_1(z)}{\sum_{a'} e^2_{a'} f^{a'}_1(x) D^{\Lambda,a'}_1(z)} dx.
\end{equation}
In Deep Inelastic Scattering (DIS), the scattering is predominantly off $u$ and $d$ quarks so $D^{\Lambda}_{LL'}$ is a measure of the light quark spin contribution to the $\Lambda$ in this case.  Furthermore, because of isospin symmetry, we should na{\"i}vely expect that $D^{\Lambda,u}_{LL'}=D^{\Lambda,d}_{LL'}$~\cite{Airapetian_2006}.

% \subsection{Previous Work}\label{sec:introduction:previous_work}

In the Na{\"i}ve Quark Model (NQM), the $s$ quark carries the entirety of the $\Lambda$ spin while the $u$ and $d$ quarks are in a spin singlet state.  Using SU(3) symmetry considerations to extrapolate measurements of the proton spin structure for the strange baryon octet, Burkardt and Jaffe found indications of a small negative light quark polarization in the $\Lambda$~\cite{PhysRevLett.70.2537}.  Furthermore, phenomenological models have been applied to DIS data to examine the $x$-dependence of the longitudinal spin transfer, which is predicted to rise with $x$~\cite{PhysRevD.61.014007,PhysRevD.65.034004,PhysRevD.62.094001}.

The longitudinal spin transfer coefficient has been measured previously at OPAL and ALEPH in $e^{+}e^{-}$ annihilation near the $Z^0$ resonance~\cite{Ackerstaff_1998,BUSKULIC1996319}.  These observations revealed a large negative longitudinal spin transfer coefficient.  However, this is associated with the strange quark content of the $\Lambda$ since the typical production mechanism is $e^{+}e^{-} \rightarrow Z^0 \rightarrow s\bar{s}$.  The longitudinal spin transfer has also been measured recently in Drell-Yan polarized $pp$ collisions at STAR where it was observed to be small and compatible with zero~\cite{PhysRevD.109.012004,PhysRevD.98.112009}.  In the Drell-Yan case, the longitudinal spin transfer coefficient is also sensitive to the strange quark spin contribution in the $\Lambda$.  In exclusive DIS where all final state particles are identified, the longitudinal spin transfer has been measured in $\Lambda K^+$ and $\Sigma K^+$ photoproduction~\cite{Bradford_2007} and electroproduction~\cite{PhysRevLett.90.131804,PhysRevC.79.065205} with the CLAS detector and more recently in electroproduction with the CLAS12 detector~\cite{PhysRevC.105.065201}.  In this context, the spin transfer coefficient offers insight into nucleon resonances in the strange quark $s$-channel.

In terms of directly comparable measurements from SIDIS, the coefficient has been measured at HERMES~\cite{Airapetian_2006} and COMPASS~\cite{KANG2007106,Kang:2007bja,Alekseev_2009} in $e^{+}$ and $\mu^{+}$ SIDIS respectively, as well as at NOMAD~\cite{ASTIER20003,Naumov_2001} in $\nu_{\mu}$ charged-current interactions.  HERMES and COMPASS both observed small longitudinal spin transfer coefficients in the CFR compatible with $0$.  NOMAD measurements were concentrated in the TFR where they observed a negative spin transfer, yet in the CFR their measured spin transfer was compatible with zero.

%% CLAS12
\subsection{The CLAS12 Detector} \label{sec:The CLAS12 Detector}
The Continuous Electron Beam Accelerator Facility (CEBAF) at Jefferson Lab delivers a high luminosity polarized electron beam to four experimental halls for fixed target experiments~\cite{doi:10.1146/annurev.nucl.51.101701.132327}.  The CLAS12 (CEBAF Large Acceptance Spectrometer for operation at $12$~GeV beam energy) detector is located in experimental Hall B and provides excellent momentum and angular coverage, as well as good particle identification capabilities for both charged and neutral particles produced in high energy electron-nucleus scattering events~\cite{BURKERT2020163419}.  The detector is centered around two large superconducting magnets, a solenoid in the central region of the detector and a torus in the forward region.  The torus magnet is operated in two different configurations, either bending negatively charged particles in toward the beamline (inbending configuration), or out away from the beamline (outbending configuration).  The detector is also separated into two major detector systems.  The Forward Detector (FD) covers the polar angle region $5\degree < \theta < 35\degree$ and the Central Detector (CD) covers the polar angle region $35\degree < \theta < 125\degree$.  Both detector regions have symmetric azimuthal coverage about the electron beam direction.

% Data
\section{Data} \label{sec:Data}
The data used in this study were all taken during the fall 2018 run period in the outbending torus field configuration with a $10.6$~GeV longitudinally polarized electron beam and a $5$~cm long unpolarized liquid-hydrogen target.  This sample corresponds to a total of roughly $35.8$~mC accumulated charge, which corresponds to an integrated luminosity of $49$~fb$^{-1}$.  The outbending configuration is necessary to identify $\Lambda$ baryons because the decay $\pi^{-}$ typically has much lower momentum than the $p$.  Thus, in the inbending configuration the $\pi^{-}$ is bent into the beamline, drastically reducing the acceptance for the $\Lambda \rightarrow p\pi^{-}$ channel.  Before any kinematic cuts were applied to the dataset, a momentum correction~\cite{capobianco23} for $e^{-}$ and $\pi^-$ was first applied to account for biases in the detector reconstruction algorithm.  The correction was developed from exclusive events where all final state particles are known and the correct momentum may be inferred from momentum conservation.  Our data were all required to have an identified scattered electron ($e^{-}$) and a proton-pion ($p\pi^{-}$) pair in the reconstruction.  The electron production vertex was required to be reconstructed within a $12.5$~cm interval within the scattering chamber excluding the scattering chamber exit window but allowing for some spread in the reconstructed value due to imperfections in alignment. The scattered electron was also required to have a momentum $p_{e^{-}}>2$~GeV.  The proton and pion were both also required to be detected in the FD ($\theta<35\degree$).

Our sample of Monte Carlo (MC) simulation events was produced with the same run configuration as the actual data from the fall 2018 run period.  Events were generated with an MC algorithm based on the PEPSI (Polarized Electron Proton Scattering Interactions) Lund program~\cite{MANKIEWICZ1992305}, which computes the photon-parton hard scattering to first order in the strong coupling constant $\alpha_s$ and uses the JETSET~\cite{SJOSTRAND1987367} routines to simulate the hadronic fragmentation.  While the statistics of the MC sample are significantly lower than that of the data sample, this is not considered an issue because there is such a preponderance of data and the MC only enters the results through fractional corrections and in the systematic uncertainties, where it is one of the smallest contributions.

The kinematic cuts to select SIDIS events for our analysis were $Q^2>1$~GeV$^2$, $W>2$~GeV, $p_{e^{-}}>2$~GeV, $y<0.8$, $z_{p\pi^{-}}<1$.  We also required $x_{F p\pi^{-}}>0$ for the $p\pi^{-}$ pair to cut out backward-traveling particles in the $\gamma^*N$ CM frame and reduce TFR contributions.  Additionally, the invariant mass of the $p\pi^{-}$ pair was restricted to $M_{p\pi^{-}}<1.24$~GeV so that events were sufficiently close to the $\Lambda$ mass peak.  After cuts, our dataset contains roughly $3.4 \times 10^6$ events.  While previous experiments~\cite{Airapetian_2006,KANG2007106,Kang:2007bja} have relied on looking for the extended $\Lambda$ decay vertex to isolate the $\Lambda$ signal and reduce combinatorial background, we did not apply such an approach in our final analysis.  This could be investigated in future analyses of this sort.  Some effort was made to use Armenteros-Podolanski plots~\cite{armenteros_1954} to isolate the $\Lambda$ signal, however, cuts based on this method were found to make the invariant mass distribution too complex to perform a reliable fit of signal and background distributions.

%% Signal Extraction
\subsection{Signal Extraction} \label{Signal Extraction}
\begin{figure}[ht!]
\centering
\includegraphics[width=0.48\textwidth]{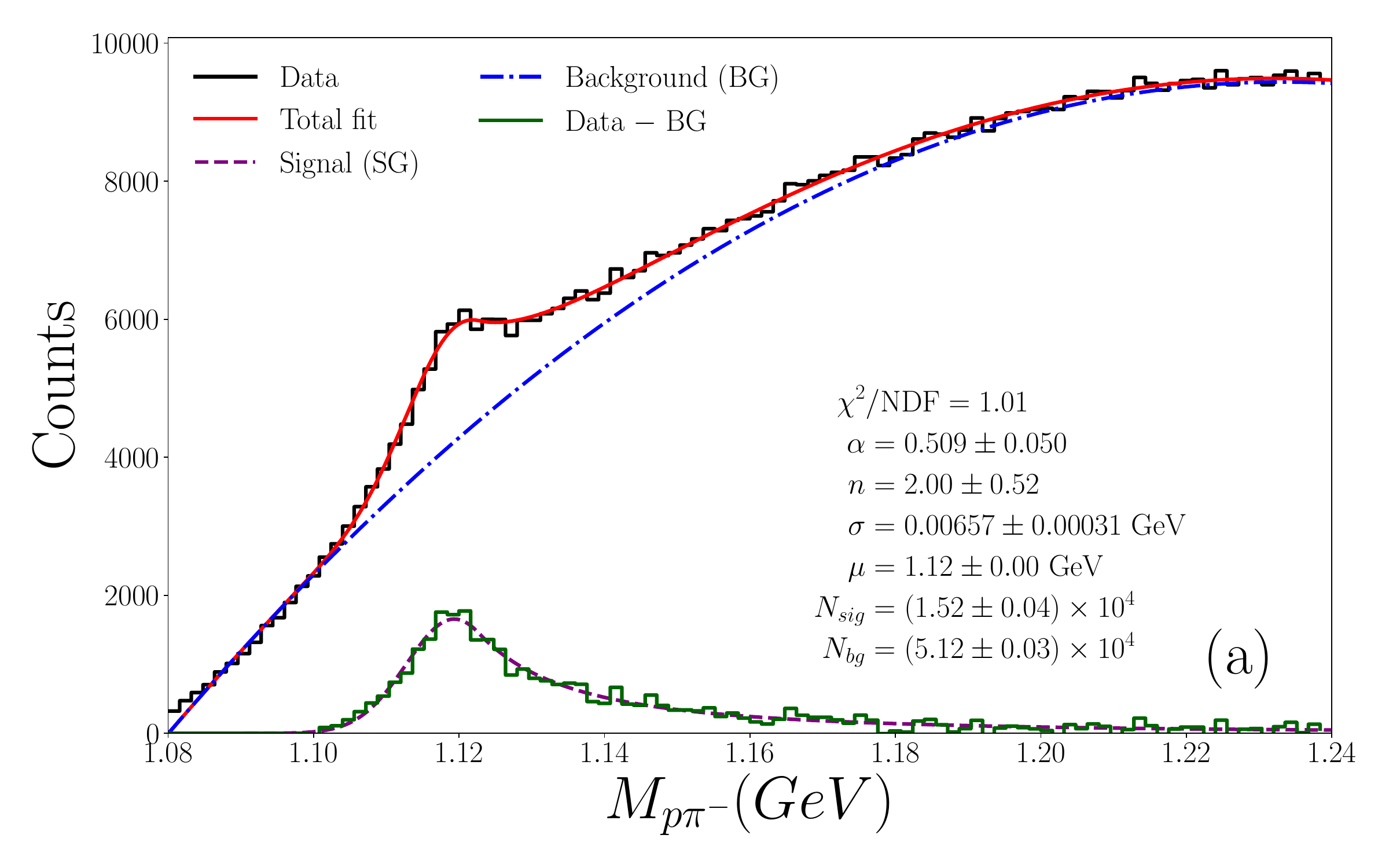}
\includegraphics[width=0.48\textwidth]{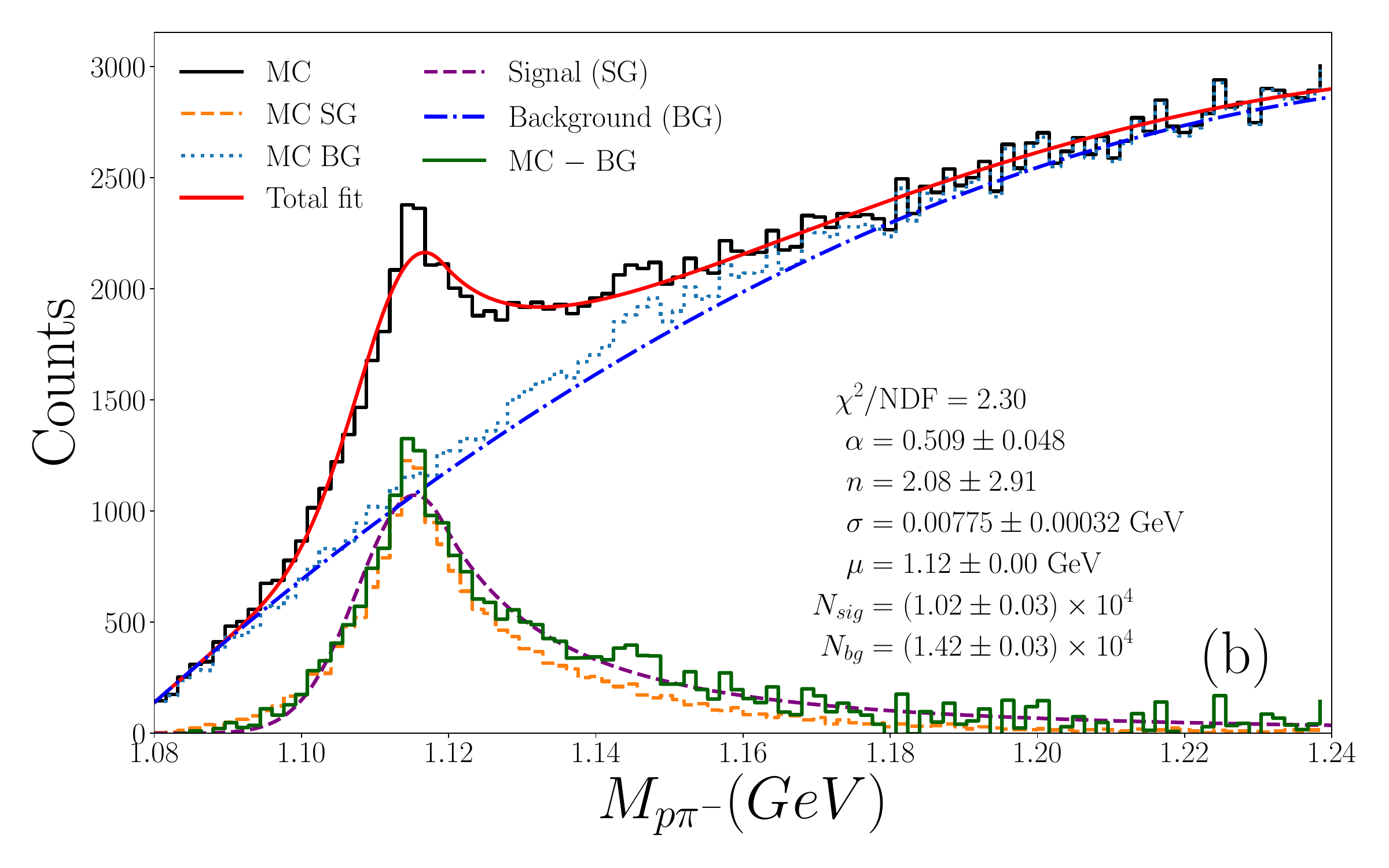}
\caption{Fits to the invariant mass spectrum $M_{p\pi^{-}}$ of reconstructed $p\pi^{-}$ pairs in data (a) and MC (b) are similar up to scale factors.  These distributions are from the middle bin $3$ in our $z_{p\pi^{-}}$ bin scheme.  The histogram counts over the fitted background are shown in the black histograms at the bottom of each plot for comparison with the fitted signal function.}
\label{fig:mass}
\end{figure}

The invariant mass spectrum of reconstructed $p\pi^{-}$ pairs passing kinematic cuts is shown in Fig.~\ref{fig:mass} for both the data and MC simulation.  A peak around the nominal $\Lambda$ mass $M=1.1157$~GeV is apparent, but the background contribution is very high, especially in data.  The main background contribution comes from combinatorics of $p\pi^{-}$ pairs that do not originate from a $\Lambda$ decay.  A Crystal Ball function~\cite{PhysRevD.34.711} was used to fit the signal in order to accurately capture the tail of the signal distribution towards higher $M_{p\pi^{-}}$ events, while the background spectrum was modeled with a simple quadratic function or a $4^{th}$-order polynomial depending on the bin.  The Crystal Ball function $f$ is defined such that both the function and its first derivative are continuous.
\begin{equation}
    f(x;\alpha,n,\sigma,\mu) = 
    N \bigg{\{}
    \begin{array}{lr}
        \text{exp}(-\frac{(x-\mu)^2}{2\sigma^2}), & \text{ } \frac{x-\mu}{\sigma} \leq \alpha\\
        A (B - \frac{x-\mu}{\sigma})^{-n}, & \text{ } \frac{x-\mu}{\sigma} > \alpha
    \end{array},
    \label{eq:methods_cb}
\end{equation}
where $N$, $A$, and $B$ are constants that are defined by the function parameters and satisfy the continuity requirements and normalization.  Before settling on the Crystal Ball signal shape, we experimented with several other shapes including a simple Gaussian and a sum and a convolution of Gaussian and Crystal Ball functions.  While not perfect, as we will discuss in the following paragraph, the Crystal Ball function provided the best option in terms of fit stability and minimum $\chi^2$ values across our kinematic bins.

The signal is more pronounced in MC but the fit parameters are consistent up to scale factors between MC and data.  Since the yields from the MC do not enter directly into the final results, the difference in relative signal and background yields between MC and data should not affect our measurements.  However, as one may notice, the peak of the signal distribution exceeds the Gaussian peak of the Crystal Ball function in both MC and data.  Thus, to minimize the effect of this excess peak, rather than integrating the fitted signal distribution, we estimated the signal events $N_{sig}$ by subtracting from the counts of the total spectrum the counts $N_{bg}$ of the integrated background function in the signal region.  One may also observe that the Crystal Ball signal shape is broad, while we know the raw $\Lambda$ spectrum before detector acceptance is extremely narrow since the $\Lambda$ has a relatively long lifetime of $(2.617\pm0.0010)\times 10^{-10}$~s~\cite{ParticleDataGroup:2024cfk}.  This is due to the poor reconstruction of the $\pi^{-}$ for the $\Lambda \rightarrow p\pi^{-}$ channel because it tends to have low momentum.  For the final results, the signal region limits were optimized so as to maximize the signal purity and minimize the signal uncertainty, which gave a signal region of $\pm2\sigma$ centered about the signal mean $\mu$ at roughly the nominal $\Lambda$ mass.  This gave us a signal region of roughly $1.11-1.13$~GeV.
%%% Lambda Mass Spectra Data and MC %%%

% Methods
\section{Methods} \label{sec:Methods}
%% Extraction of D_LL
\subsection{Extraction of \texorpdfstring{$D^{\Lambda}_{LL'}$}{DLL}} \label{Extraction of D_LL}
To extract the spin transfer coefficient $D^{\Lambda}_{LL'}$, one may use the na{\"i}ve method of simply fitting the acceptance-corrected $\cos{\theta_{pL'}}$ distribution to a linear function and using the fitted slope to compute $D^{\Lambda}_{LL'}$. However, this method we found to systematically undershoot injected asymmetries, and hence, for our analysis we used a slightly more sophisticated method that we refer to as the Helicity Balance (HB) method.  A detailed derivation of this method may be found in Ref.~\cite{Schnell_1999}.  It is based on the method of maximum likelihood and relies on the assumption that the luminosity averaged beam polarization $\overline{\lambda_{\ell}}$ is consistent with zero.  This allows the acceptance function to cancel out in the derivation since it couples to $\overline{\lambda_{\ell}}$ for computing quantities that are linearly dependent on $\lambda_{\ell}$.  Following this method, one may calculate the spin transfer coefficient with a simple numerical formula
\begin{equation}
    D^{\Lambda}_{LL'} = \frac{1}{\alpha_{\Lambda} \overline{\lambda_{\ell}^2}}\frac{\sum^{N_{\Lambda}}_{i=1}\lambda_{\ell,i} \cos{\theta_{pL'}^i}}{\sum^{N_{\Lambda}}_{i=1}D(y_i) \cos^2{\theta_{pL'}^i}},
\label{eq:methods_hb_final}
\end{equation}
where $\overline{\lambda_{\ell}^2}$ is the luminosity-averaged squared beam polarization, which was $(89.22\pm2.51)\%$ for our dataset.  The sums run over the events in a given kinematic bin.

Of course, this method has its limitations.  We observed this by injecting an artificial asymmetry in MC simulation to see the difference between the true, injected value and the extracted value of the asymmetry.  We applied a correction in each kinematic bin from the fractional difference of the extracted to the injected asymmetry in MC scaled to the asymmetry extracted in data.  The corrections are computed as follows.  We first take the difference between reconstructed and injected asymmetry $\Delta A_{MC} = A_{Rec}-A_{Inj}$ in MC for each data point.  We then scale this difference to the measured asymmetry to obtain the correction $\Delta A_{Data}=\Delta A_{MC} \cdot \frac{A_{Data}}{A_{MC}^{Inj}}$.  The corrected value is then $A_{Data}^{Cor}=A_{Data}-\Delta A_{Data}$.  The underlying assumption, which generally holds true for asymmetries, is that the extracted asymmetry should scale proportionally to the true asymmetry.  The asymmetry we injected was $0.1$ for these corrections since this was on par with the average asymmetry.  The average value of the corrections was $+23\%$ for $\cos{\theta_{LL'}}$ along $P_{\Lambda}$ and $+25\%$ for $\cos{\theta_{LL'}}$ along $P_{\gamma^{*}}$.  The corrections have a non-trivial kinematic dependence which is why we apply them independently in each bin.

%% Background Correction
\subsection{Background Correction} \label{Background Correction}
Since there is still an irreducible background underneath our $\Lambda$ signal due to combinatorics of particles not originating from the same $\Lambda$ decay or any $\Lambda$ decay at all, we have to correct our results to subtract away any background contributions.  To do this we use the method of sideband subtraction.  Assuming that the background polarization does not vary much as a function of the invariant mass $M_{p\pi^{-}}$ we can calculate the polarization in the signal region and in a sideband region away from the signal.  

Since we can compute the relative background fraction $\varepsilon$ in the signal region from our signal fit we can then use a properly weighted subtraction of the sideband polarization to correct for any background polarization in the signal region.
\begin{equation}
    D_{LL'}^{\Lambda} = \frac{D_{LL'sig} - \varepsilon D_{LL'sb}}{1 - \varepsilon},
\label{eq:sideband_subtraction}
\end{equation}
where $D_{LL'sb}$ is the background contribution from the sideband region and $D_{LL'sig}$ is the contribution in the signal region.  For our purposes we define our signal region as the $\pm2\sigma$ window about the signal mean $\mu$, where $\sigma$ is the signal width obtained from the fit to the invariant mass spectrum.  For the unbinned mass spectrum this region is $1.11$~GeV$<M_{p\pi^{-}}<1.13$~GeV.  The sideband regions for this analysis were chosen to be $1.08$~GeV$<M_{p\pi^{-}}<1.10$~GeV and $1.15$~GeV$<M_{p\pi^{-}}<1.18$~GeV.  We use the same signal and sideband regions across all bins.

\begin{figure}[ht] % ht!]
\includegraphics[width=0.48\textwidth, left]{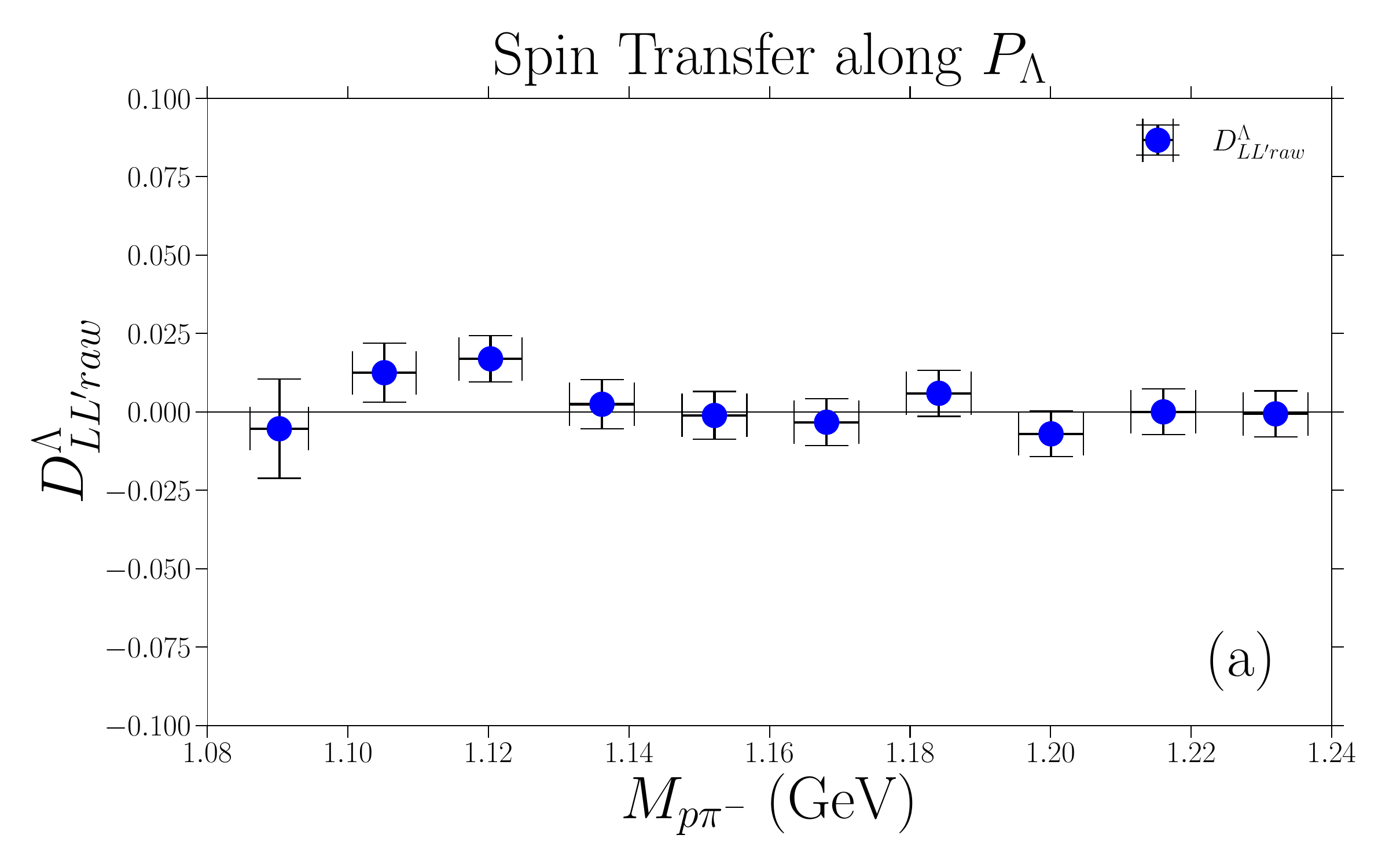}
\includegraphics[width=0.48\textwidth, left]{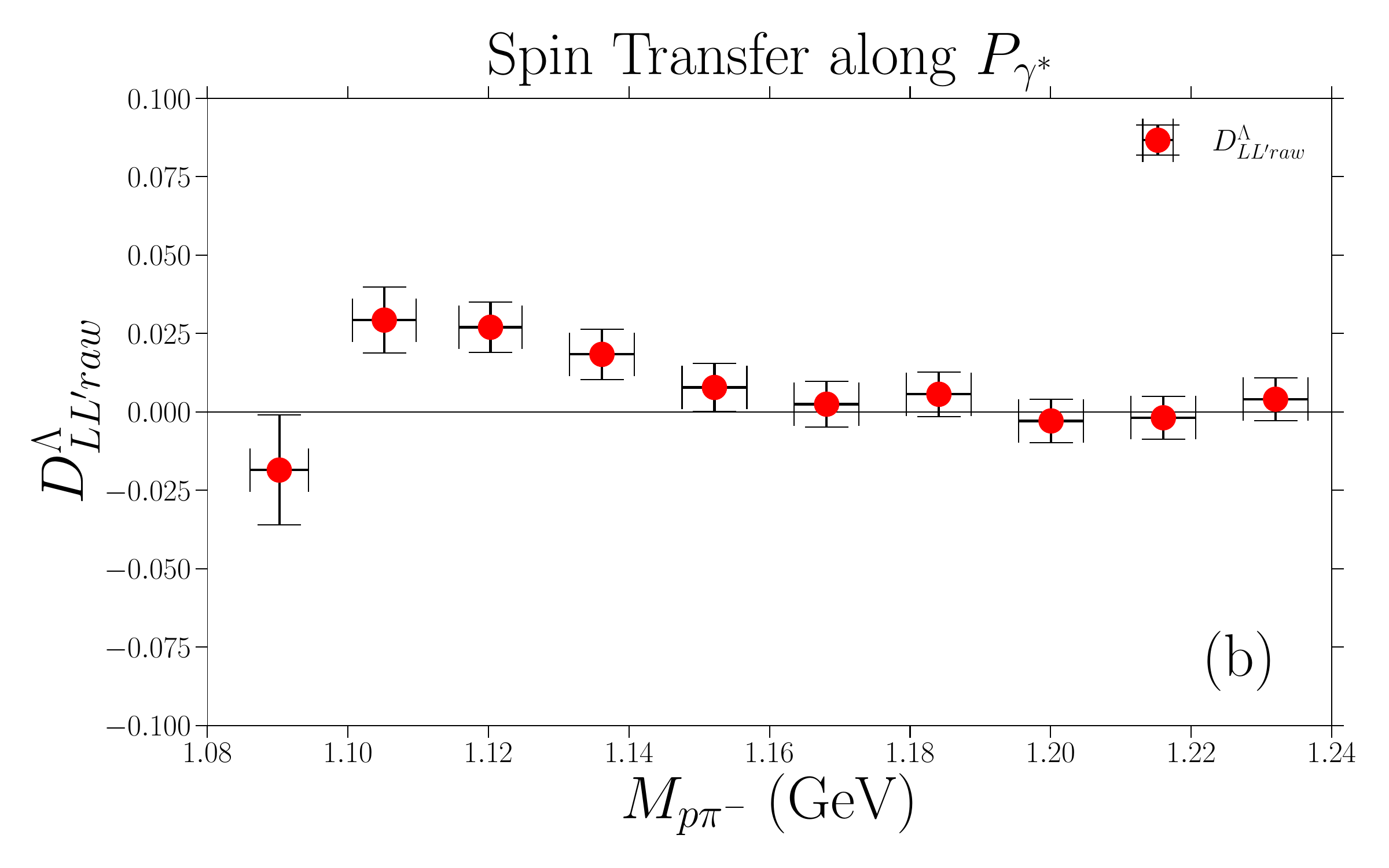}
\caption{Raw spin transfer results obtained with the HB method, \textbf{before} background correction, as a function of the mass spectrum $M_{p\pi^{-}}$ over the entire dataset.  The subpanels show results for $\Lambda$ polarization along the $\Lambda$ momentum (a) and along the virtual photon $\gamma^*$ momentum (b) in the $\gamma^*N$ CM frame.  Vertical uncertainties are statistical and horizontal uncertainties are given by the standard deviation within each bin.}
\label{fig:systematics_kinematics_mass_ppim}
\end{figure}

The sideband subtraction depends on the validity of the assumption that the background polarization does not strongly depend on $M_{p\pi^{-}}$ near the $\Lambda$ signal region.  Figure~\ref{fig:systematics_kinematics_mass_ppim} shows the raw polarization extracted using the HB method \textit{before} any background correction as a function of $M_{p\pi^{-}}$.  As one would hope, we see some slight dependence near the $\Lambda$ signal region and essentially a flat distribution close to zero elsewhere.
%%% FIG: Mass Dependence of Polarization HB %%%

%% Kinematic Bins
\subsection{Kinematic Bins and Coverage}
\begin{figure}[ht!]
\centering
\includegraphics[width=0.45\textwidth]{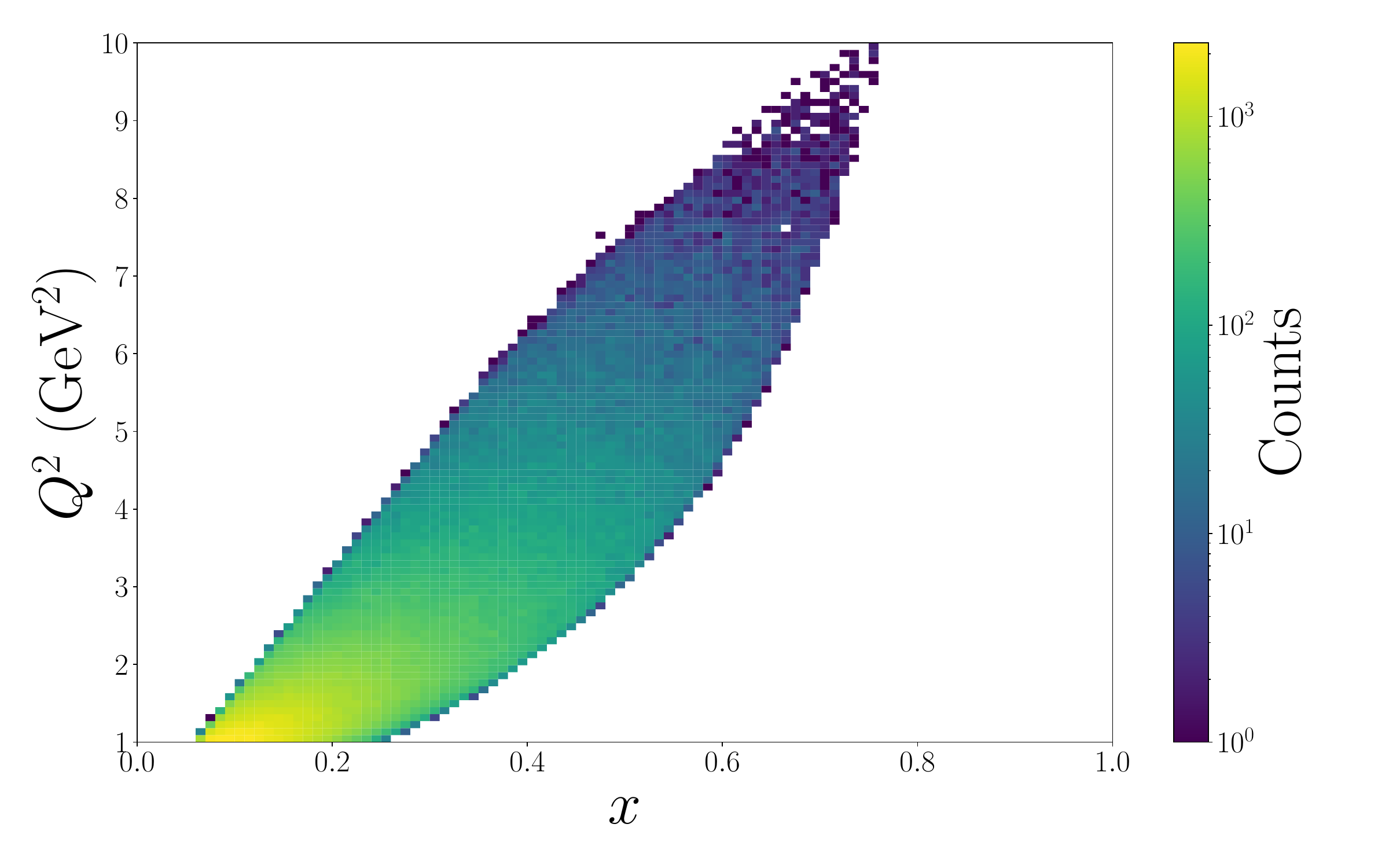}
\includegraphics[width=0.45\textwidth]{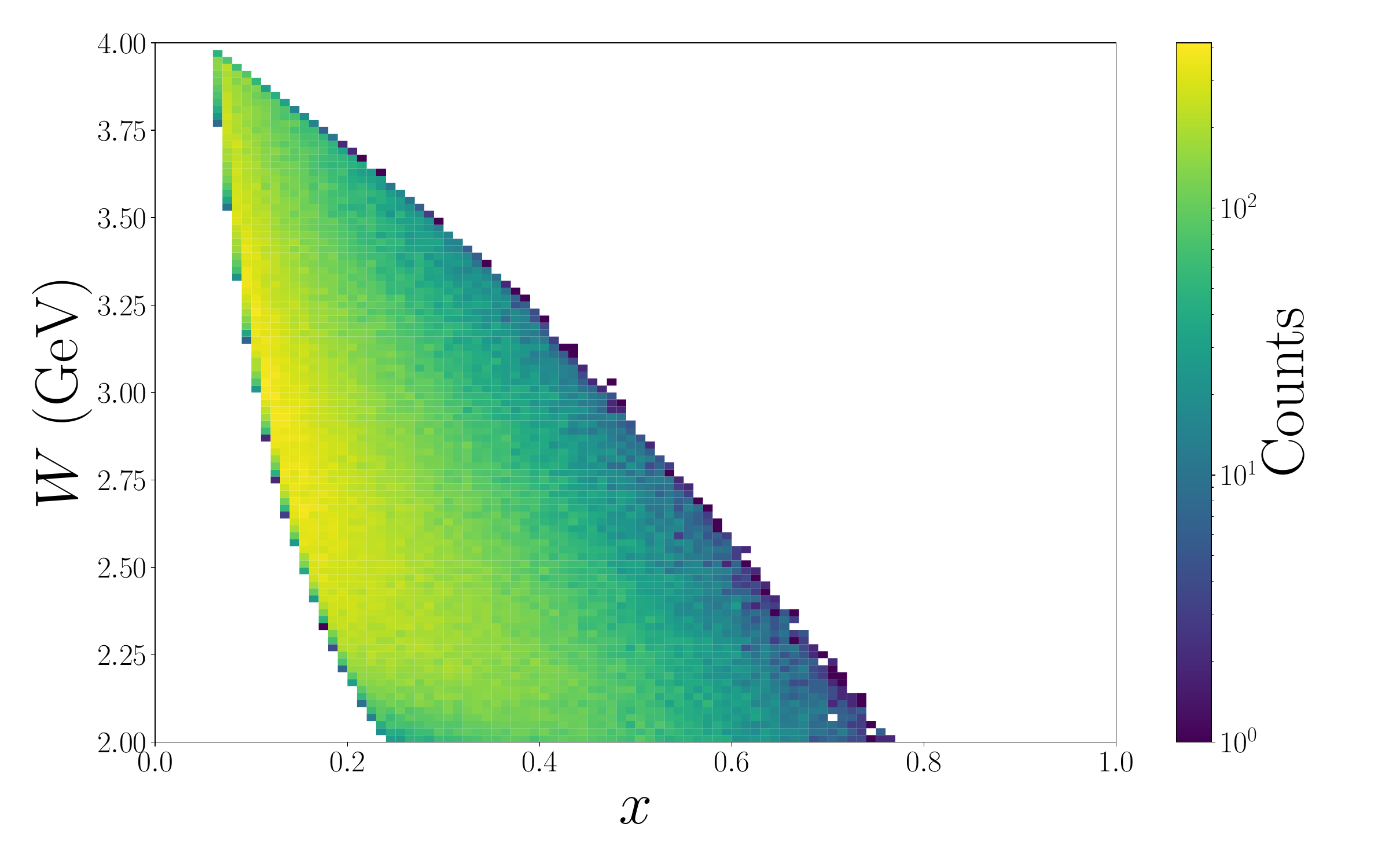}
\caption{Kinematic coverage in $Q^2$, $x$, and $W$ of our dataset.}
\label{fig:methods_dt_2d_kinematics}
\end{figure}

We divided our dataset into \textit{separate} 1D binning schemes in the kinematic variables $z_{p\pi^{-}}$ and $x_{F p\pi^{-}}$.  The bin definitions in both variables are given in Table~\ref{table:methods_binlims}.  Five bins were chosen in each bin variable to cover the range of the variable and also to maintain reasonable statistics in each bin.  The kinematic coverage of our dataset in $Q^2$, $x$, and $W$ is shown in Fig.~\ref{fig:methods_dt_2d_kinematics}.

%%% Bin limits %%%
\begin{table}[ht!]
\centering
\begin{filecontents*}{tables/methods/binlims.csv}
bin,zbincut,xFbincut
0,0.346\leq z_{p\pi^{-}}<0.593,0.000\leq x_{F p\pi^{-}}<0.050
1,0.593\leq z_{p\pi^{-}}<0.686,0.050\leq x_{F p\pi^{-}}<0.108
2,0.686\leq z_{p\pi^{-}}<0.770,0.108\leq x_{F p\pi^{-}}<0.178
3,0.770\leq z_{p\pi^{-}}<0.860,0.178\leq x_{F p\pi^{-}}<0.278
4,0.860\leq z_{p\pi^{-}}<1.000,0.278\leq x_{F p\pi^{-}}<0.805
\end{filecontents*}
\csvreader[
        head to column names,
        before reading = \begin{center}\sisetup{table-number-alignment=center},
        tabular = c || c | c ,
        table head = \textbf{Bin} & \textbf{$z_{p\pi^{-}}$ Bins} & \textbf{$x_{F p\pi^{-}}$ Bins} \\\midrule,
        table foot = \bottomrule,
        after reading = \end{center},
]{tables/methods/binlims.csv}
{}{ \bin & $\zbincut$ & $\xFbincut$}
\caption{Bin limits used for kinematically binned results.}
\label{table:methods_binlims}
\end{table}

%%% DT x 2D Correlations %%%

% %%% MC x 2D Correlations %%%
% \begin{figure}[ht!]
% \centering
% \includegraphics[width=0.45\textwidth]{figs/data/correlations_mc/c_2d_mc__x_Q2.pdf}
% \includegraphics[width=0.45\textwidth]{figs/data/correlations_mc/c_2d_mc__x_W.pdf}
% \caption{Kinematic coverage in $Q^2$, $x$, and $W$ of our MC dataset}
% \label{fig:methods_mc_2d_kinematics}
% \end{figure}

%% Bin Migration
\subsection{Bin Migration}
Due to the smearing of our reconstructed kinematic variables introduced by detector resolution, there is inevitably some migration of the reconstructed values between the true kinematic bins.  Naturally, one may unfold the distribution using the bin migration matrix $f_{i \rightarrow j}=f_{ij}^{T}$, i.e., the fraction of true $\Lambda$s generated in bin $i$ that are reconstructed in bin $j$, estimated from MC.  Assuming the measured results $A_j$ are related to the true results by the matrix equation
\begin{equation}
    A_{j} = f_{ij}^{T}A_{i,true},
    \label{eq:methods_bin_migration_transform}
\end{equation}
one can simply compute the inverse of this matrix and correct for this effect
\begin{equation}
    f_{ij}^{T-1}A_{j} = A_{i, true}.
    \label{eq:methods_bin_migration_inverse}
\end{equation}
We use this method to unfold our results, but in general the bin migration fractions are $<10\%$ and at most they are $\sim18\%$.

% Systematic Uncertainties
\section{Systematic Uncertainties} \label{Systematic Uncertainties}
We estimated our systematic uncertainties based on several sources.  A summary of average systematic values is shown in Table~\ref{tab:systematics_summary}, and breakdowns of the systematic uncertainties in each kinematic bin are shown in Appendix~\ref{app:Tables}.  We folded in the uncertainty in the $\Lambda$ decay asymmetry parameter $\delta\alpha_{\Lambda}=0.009$~\cite{ParticleDataGroup:2024cfk} as a systematic scale uncertainty.  A scale uncertainty of $2.51\%$ from our beam polarization was also included.

We estimated bin-dependent uncertainties from the variation in results due to the choice of the signal fit function for identifying the $\Lambda$ invariant mass peak by comparing the difference in results corrected with the background fraction $\varepsilon$ taken from a Gaussian signal fit and our chosen Crystal Ball signal.  In general, this was our dominant source of systematic uncertainty, and the Gaussian comparison was intentionally used as a conservative estimate of this systematic since any choice of signal shape was not an exact match to the actual signal distribution.  However, our overall systematics tend to be much smaller than the statistical uncertainties even with this conservative estimate.

We also estimated a bin-dependent systematic uncertainty on our extraction via the HB method from the variation in the deviation of results from an injected asymmetry for several values of the injected asymmetry, which we assessed using MC simulated events.  We included a final bin-dependent systematic also based on asymmetry injections of $D^{\Lambda}_{LL'}$ with an additional $\cos{\phi_{\Lambda}}$-dependent cross section term.  This additional term cancels out with integration over $P_{\Lambda\perp}$~\cite{PhysRevD.95.074026}, however detector acceptance introduces a lingering effect.  We quantified the extent of this effect based on the variation in results when several different values for the additional asymmetry associated with this modulation.  We scaled this variation to data by using the ratio between data and MC of the maximum difference in the results, i.e., the difference in results between the regions where the $\cos{\phi_{\Lambda}}$ term is positive --- $\phi_{\Lambda}\in(0,\pi/2]$ or $\phi_{\Lambda}\in(3\pi/2,2\pi]$ --- or negative --- $\phi_{\Lambda}\in(\pi/2,3\pi/2]$.  To get the final systematic uncertainty scales for each kinematic bin, we added all systematic uncertainties for the bin in quadrature assuming all systematic uncertainties were completely uncorrelated.

%%% Systematics Summary %%%
\begin{table}[ht!]
\centering
\begin{filecontents*}{tables/systematics/systematics_averages.csv}
,source,averagelambda,averagegamma
0,Decay Parameter $\alpha_{\Lambda}$,0.001,0.002
1,Beam Polarization $\overline{\lambda^2_{\ell}}$,0.003,0.005
2,MC Injection,0.007,0.011
3,Mass Fit,0.008,0.013
4,$\cos{\phi_\Lambda}$ Effects,0.022,0.024
\end{filecontents*}
\csvreader[
        head to column names,
        before reading = \begin{center}\sisetup{table-number-alignment=center},
        tabular = l c c,
        table head = \textbf{Source} & \textbf{$P_{\Lambda}$} & \textbf{$P_{\gamma^*}$} \\\midrule,
        table foot = \midrule Total & 0.0130 & 0.0211\\\bottomrule,
        after reading = \end{center},
]{tables/systematics/systematics_averages.csv}
{}{\source & \averagelambda & \averagegamma}
\caption{Systematic uncertainties by source averaged over all kinematic bins for each choice of $\Lambda$ spin axis.}
\label{tab:systematics_summary}
\end{table}

% Results
\section{Results and Discussion}

All final results were calculated with the helicity balance method and may be accessed from the CLAS Physics Database~\cite{clasdatabase}.  Results over the full dataset are listed in Table~\ref{table:results_overall} and results binned in $z_{p\pi^{-}}$ and $x_{F p\pi^{-}}$ are shown in Figs.~\ref{fig:results_z_ppim}-\ref{fig:results_xF_ppim} and listed in Appendix~\ref{app:Tables}.  Uncertainties on the kinematic variables are taken as the standard deviation of the values in a given bin.

%%% Overall results %%%
\begin{table}[ht!]
\centering
\begin{filecontents*}{tables/results/overall_mass_ppim_results_and_systematics.csv}
bin,x,y,xerr,yerr,xerrsyst,yerrsyst
,P_{\Lambda},0.071,0.00537,0.029,0,0.025
,P_{\gamma *},0.130,0.00537,0.031,0,0.031
\end{filecontents*}
\csvreader[
        head to column names,
        before reading = \begin{center}\sisetup{table-number-alignment=center},
        tabular = c c,
        table head = \textbf{$\Lambda$ Spin Axis} & \textbf{Extracted Asymmetry} \\\midrule,
        table foot = \bottomrule,
        after reading = \end{center},
]{tables/results/overall_mass_ppim_results_and_systematics.csv}
{}{ $\x$ & \y$\hspace{0.1cm}\pm$\hspace{0.1cm}\yerr\hspace{0.1cm}\text{(stat)}\hspace{0.1cm}$\pm$\hspace{0.1cm}\yerrsyst\hspace{0.1cm}\text{(syst)}}
\caption{Results for $D_{LL'}^{\Lambda}$ over the full dataset.}
\label{table:results_overall}
\end{table}

%%% z_ppim Binning %%%
\begin{figure}[ht!]
\includegraphics[width=0.48\textwidth, left]{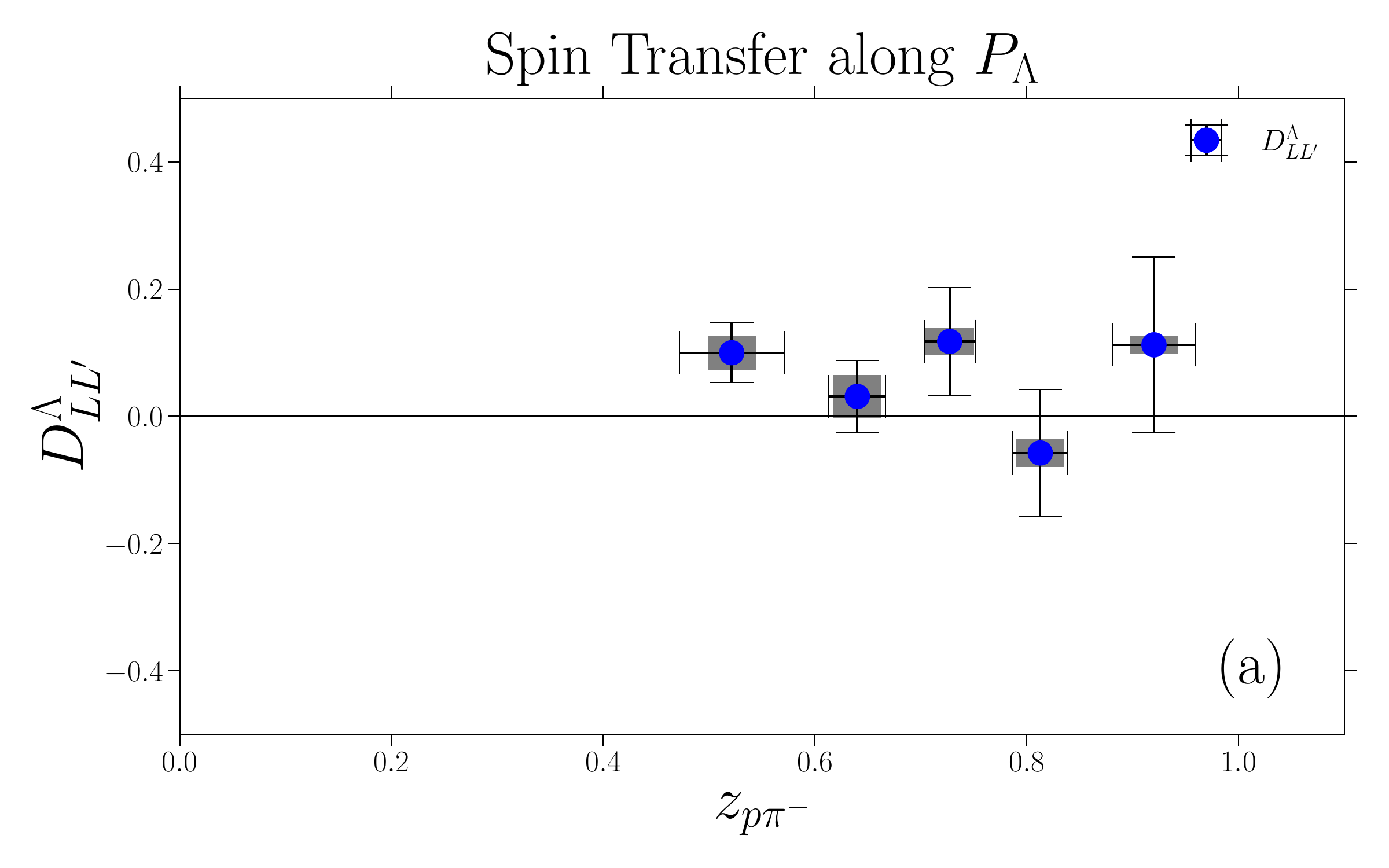}
\includegraphics[width=0.48\textwidth, left]{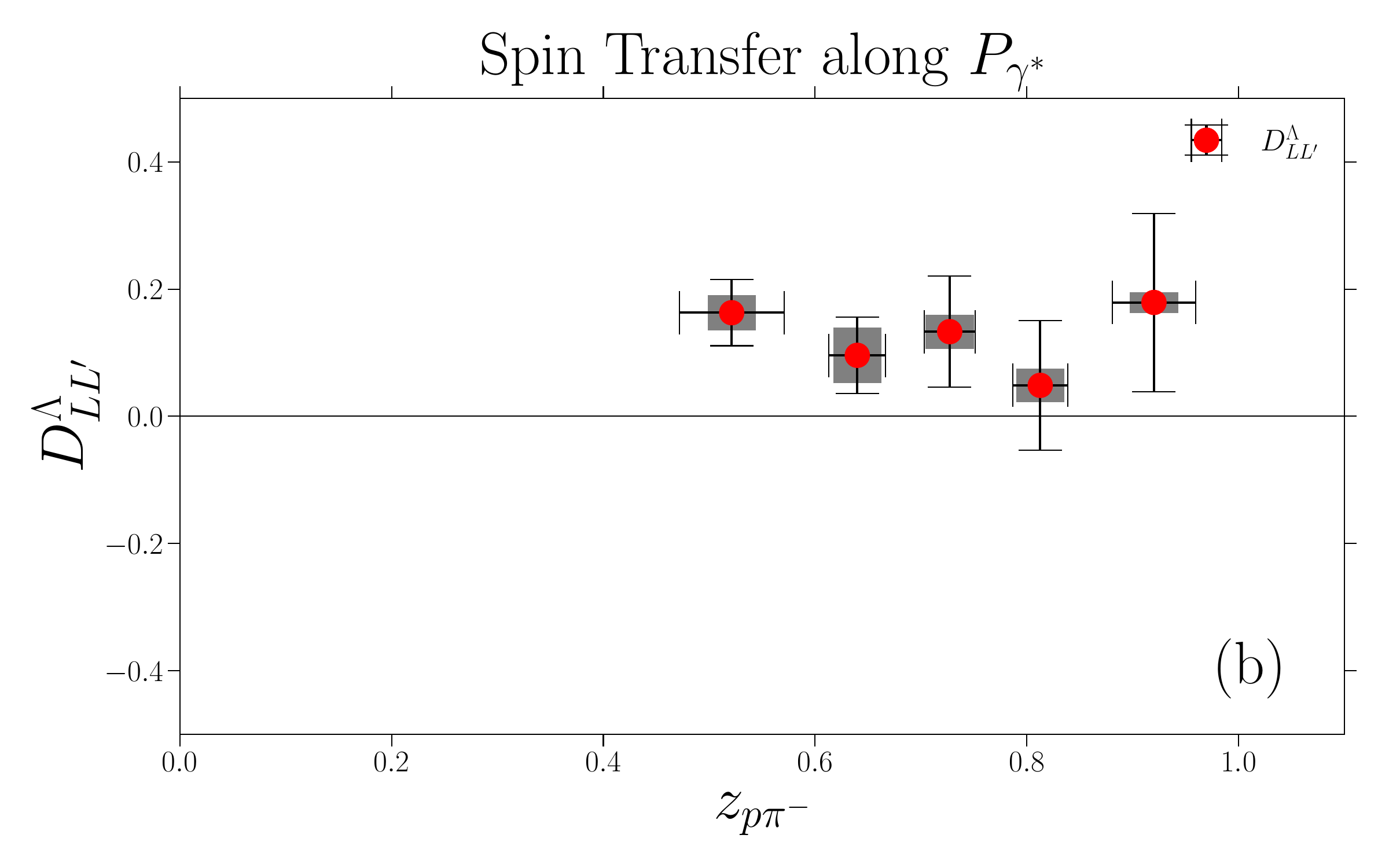}
\caption{Results for $D_{LL'}^{\Lambda}$ binned in $z_{p\pi^{-}}$.  The subpanels show results for $\Lambda$ polarization along the $\Lambda$ momentum (a) and along the virtual photon $\gamma^*$ momentum (b) in the $\gamma^*N$ CM frame.  Vertical uncertainties are statistical and horizontal uncertainties are given by the standard deviation within each bin.  Gray boxes show the systematic uncertainties.}
\label{fig:results_z_ppim}
\end{figure}

%%% xF_ppim Binning %%%
\begin{figure}[ht!]
\includegraphics[width=0.48\textwidth, left]{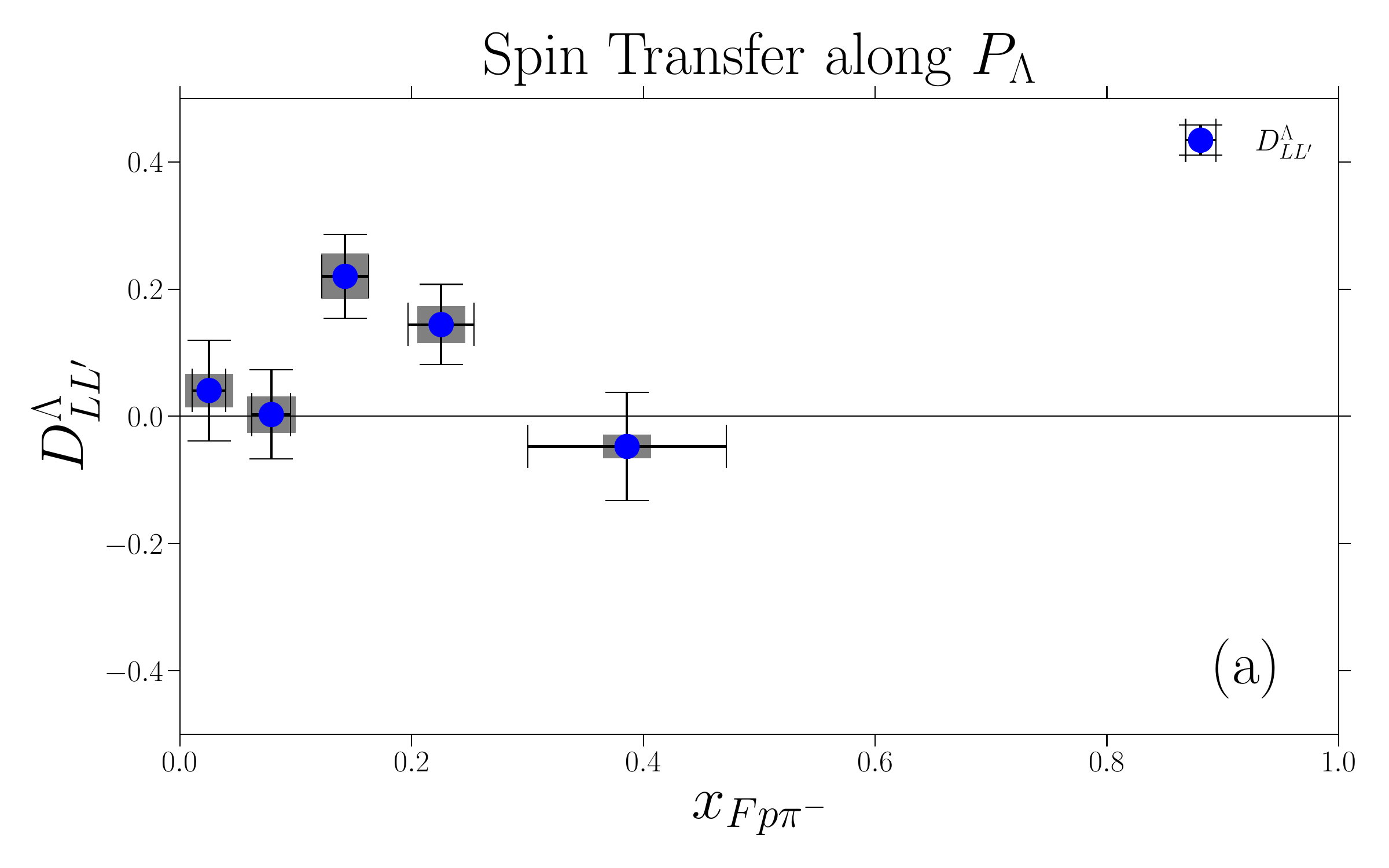}
\includegraphics[width=0.48\textwidth, left]{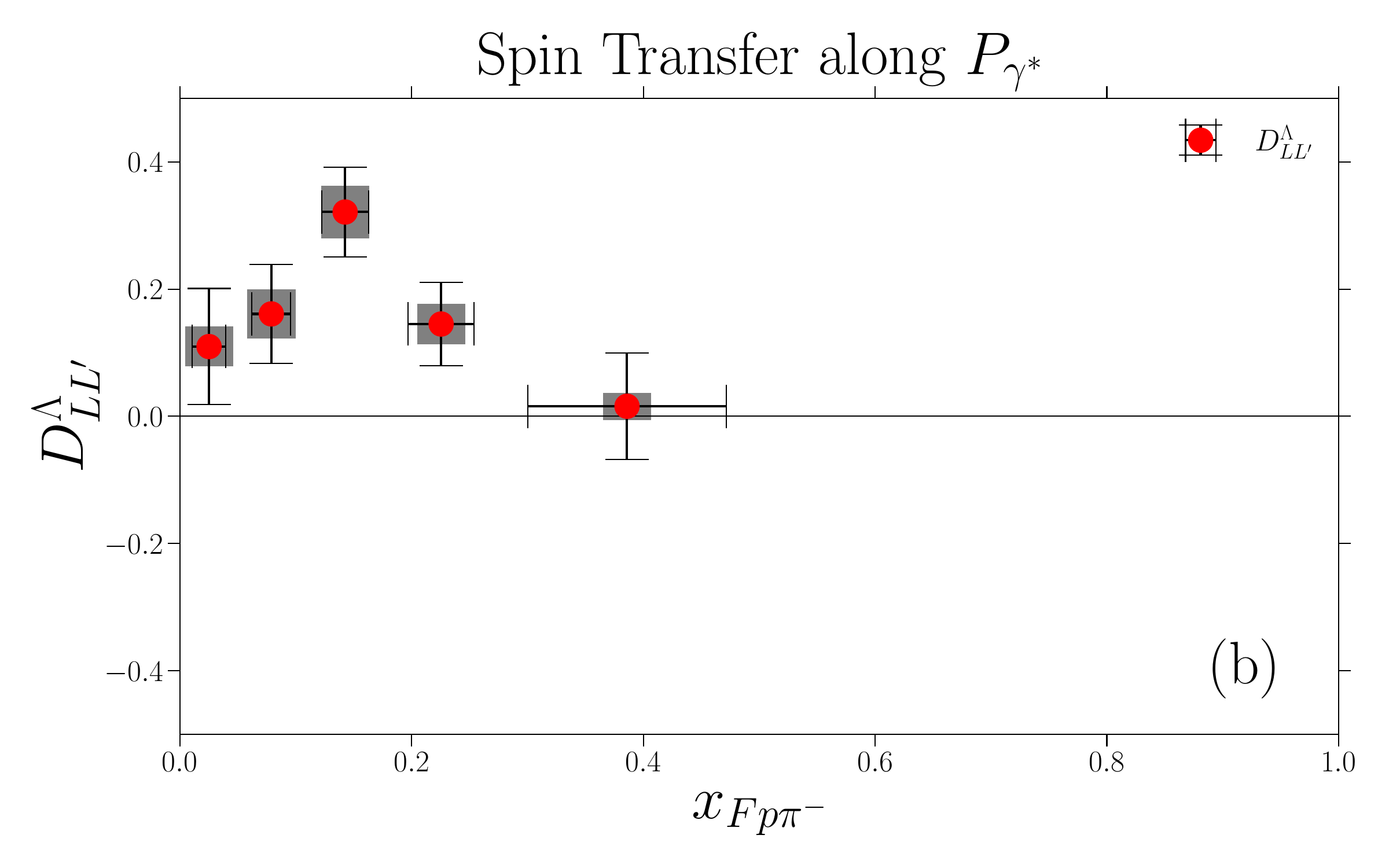}
\caption{Results for $D_{LL'}^{\Lambda}$ binned in $x_{F p\pi^{-}}$.  The subpanels show results for $\Lambda$ polarization along the $\Lambda$ momentum (a) and along the virtual photon $\gamma^*$ momentum (b) in the $\gamma^*N$ CM frame.  Vertical uncertainties are statistical and horizontal uncertainties are given by the standard deviation within each bin.  Gray boxes show the systematic uncertainties.}
\label{fig:results_xF_ppim}
\end{figure}

\begin{figure}[ht!]
\includegraphics[width=0.48\textwidth, left]{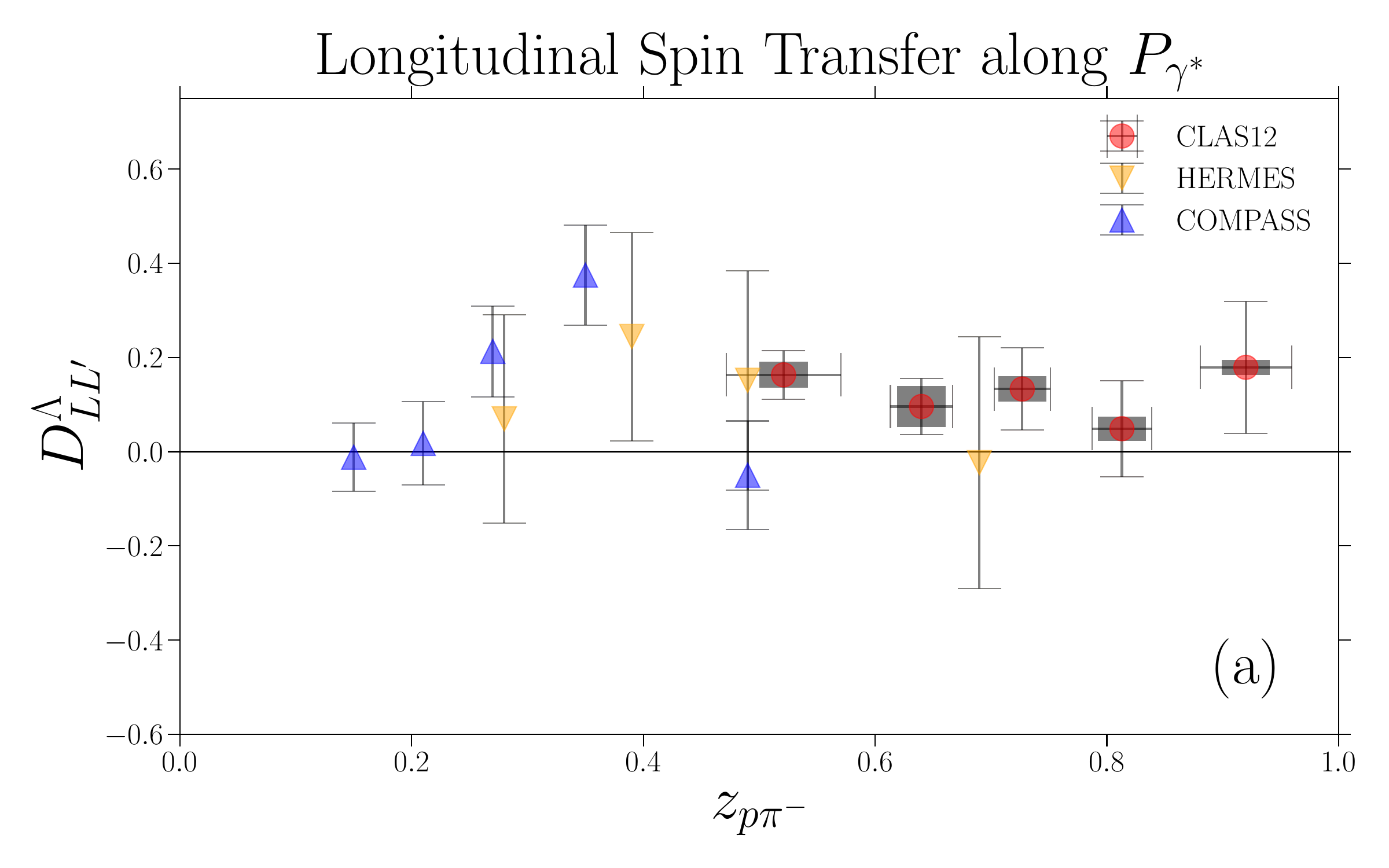}
\includegraphics[width=0.48\textwidth, left]{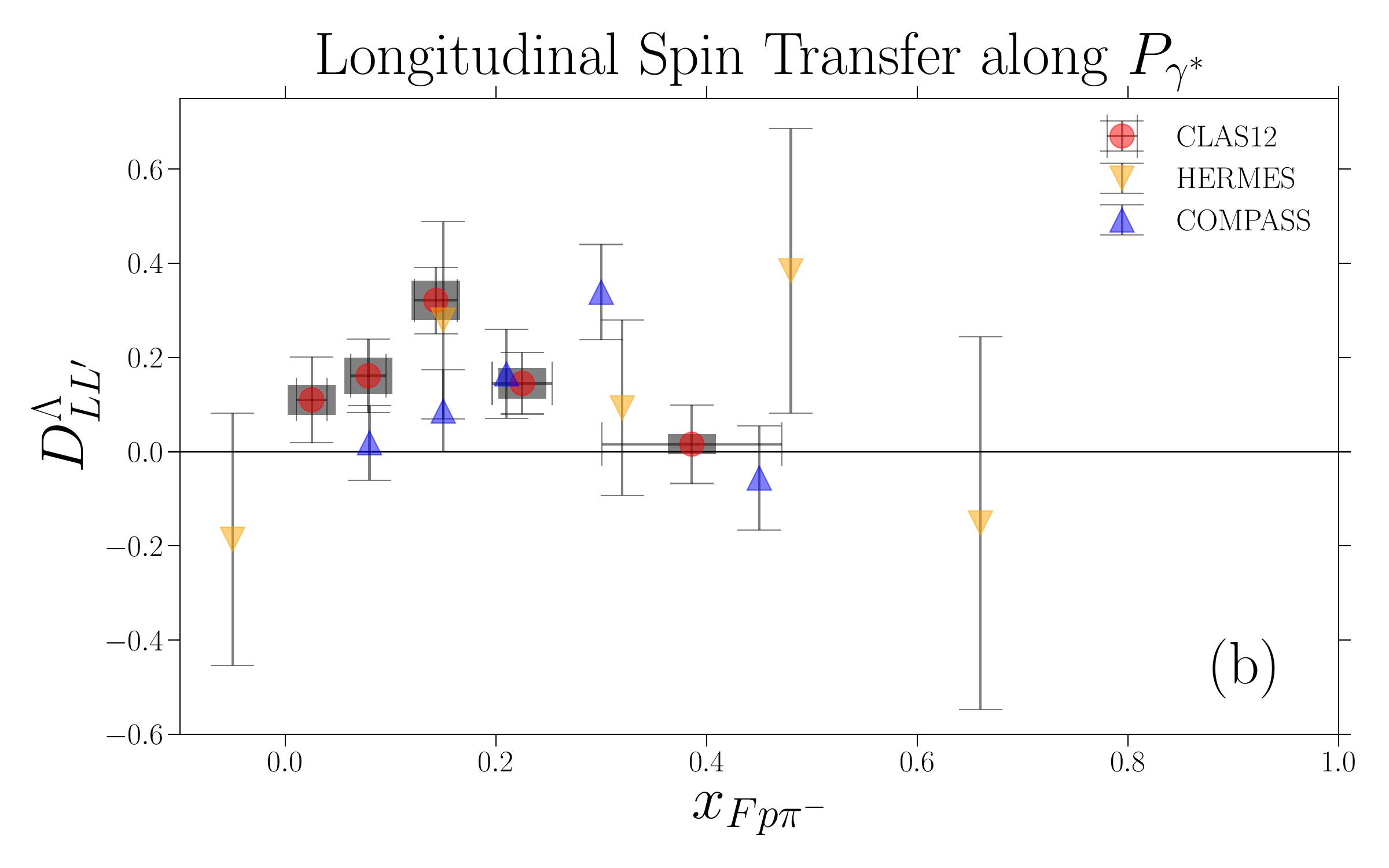}
\caption{Comparison of the extracted $D^{\Lambda}_{LL'}$ along $P_{\gamma^{*}}$ with previous results from HERMES~\cite{Airapetian_2006} and COMPASS~\cite{Kang:2007bja}.  The subpanels show results binned in $z_{p\pi^{-}}$ (a) and $x_{F p\pi^{-}}$ (b) for $\Lambda$ polarization along the virtual photon momentum in the $\gamma^*N$ CM frame.  Vertical uncertainties are statistical and horizontal uncertainties are given by the standard deviation within each bin.  Gray boxes show the systematic uncertainties.}
\label{fig:conclusions_previous_experiments}
\end{figure}

We have extracted the longitudinal spin transfer $D_{LL'}^{\Lambda}$ to $\Lambda$ hyperons in the CFR with SIDIS events at CLAS12.  In general, our results indicate a small positive light quark polarization in the $\Lambda$ hyperon.  This is as one would expect for $\Lambda$s coming primarily from $u$ and $d$ quark fragmentation.  Our results are consistent with and improve on the statistical uncertainties of previous measurements from HERMES~\cite{Airapetian_2006}, COMPASS~\cite{KANG2007106,Kang:2007bja}, and NOMAD~\cite{ASTIER20003,Naumov_2001}.  A comparison of some of these previous measurements with the results of this work is shown in Fig.~\ref{fig:conclusions_previous_experiments}.  In general, our results are also similar between the two different choices of polarization axis, as one would expect since $\Lambda$s from the CFR should carry away most of the momentum of the $\gamma^*$ considering the modest beam energy of our dataset. %Furthermore, our results show a slight positive increase with increasing $z_{p\pi^{-}}$ at high $z_{p\pi^{-}}$ which is consistent with some phenomenological predictions~\cite{PhysRevD.61.014007,PhysRevD.65.034004,PhysRevD.62.094001}.

%%% Fig: Comparison of results with previous experiments %%%

\begin{figure}[ht!]
\centering
\includegraphics[width=0.38\textwidth]{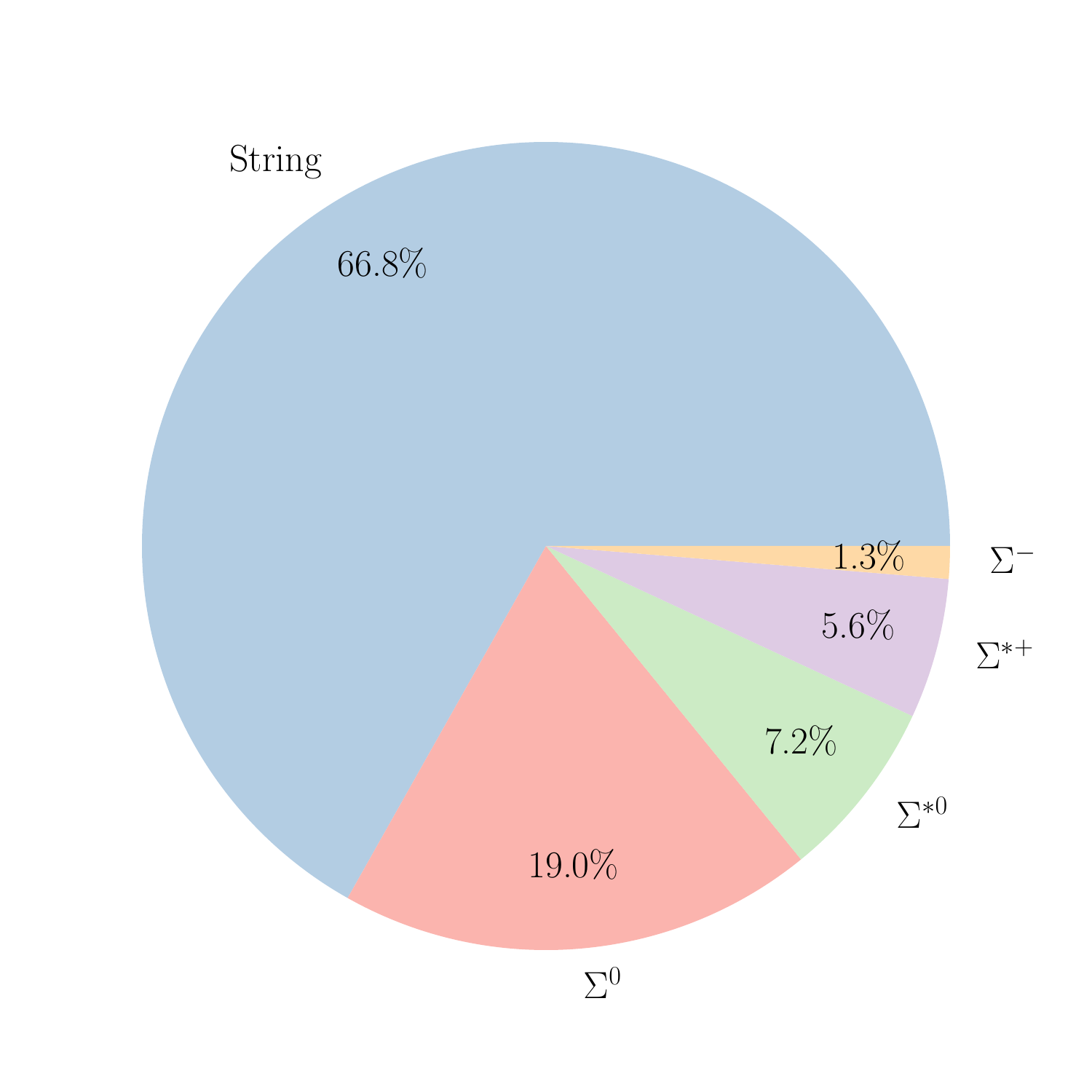}
\caption{Distribution of $\Lambda$ parents from MC~\cite{MANKIEWICZ1992305} for $\Lambda$s in the signal region with $x_{F p\pi^{-}}>0$.}
\label{fig:conclusions_lambda_parents}
\end{figure}

One must also consider that our $\Lambda$ sample contains some $\Lambda$s produced from the decays of heavier hyperons.  Our MC simulation sample shows that these are primarily from $\Sigma$ baryons, for example from the channel $\Sigma^0 \rightarrow \Lambda\gamma$ as shown in Fig.~\ref{fig:conclusions_lambda_parents}.  These decays account for $\simeq 33\%$ of the true $\Lambda$ baryons in our signal region for our MC events.  We expect the $\Lambda$ polarization to be $\text{P}_{\Lambda} = -\frac{1}{3}\text{P}_{\Sigma^{0}}$~\cite{PhysRev.109.610,ASHERY1999263} and for $\Sigma^{*} \rightarrow \Lambda\pi$ we expect $\text{P}_{\Lambda} = \frac{5}{3}\text{P}_{\Sigma^{*}}$~\cite{ASHERY1999263}.  However, there is yet no clean way to separate these feed-down $\Lambda$s from prompt $\Lambda$s originating directly from the struck quark in SIDIS.

\begin{figure}[ht!]
\includegraphics[width=0.48\textwidth, left]{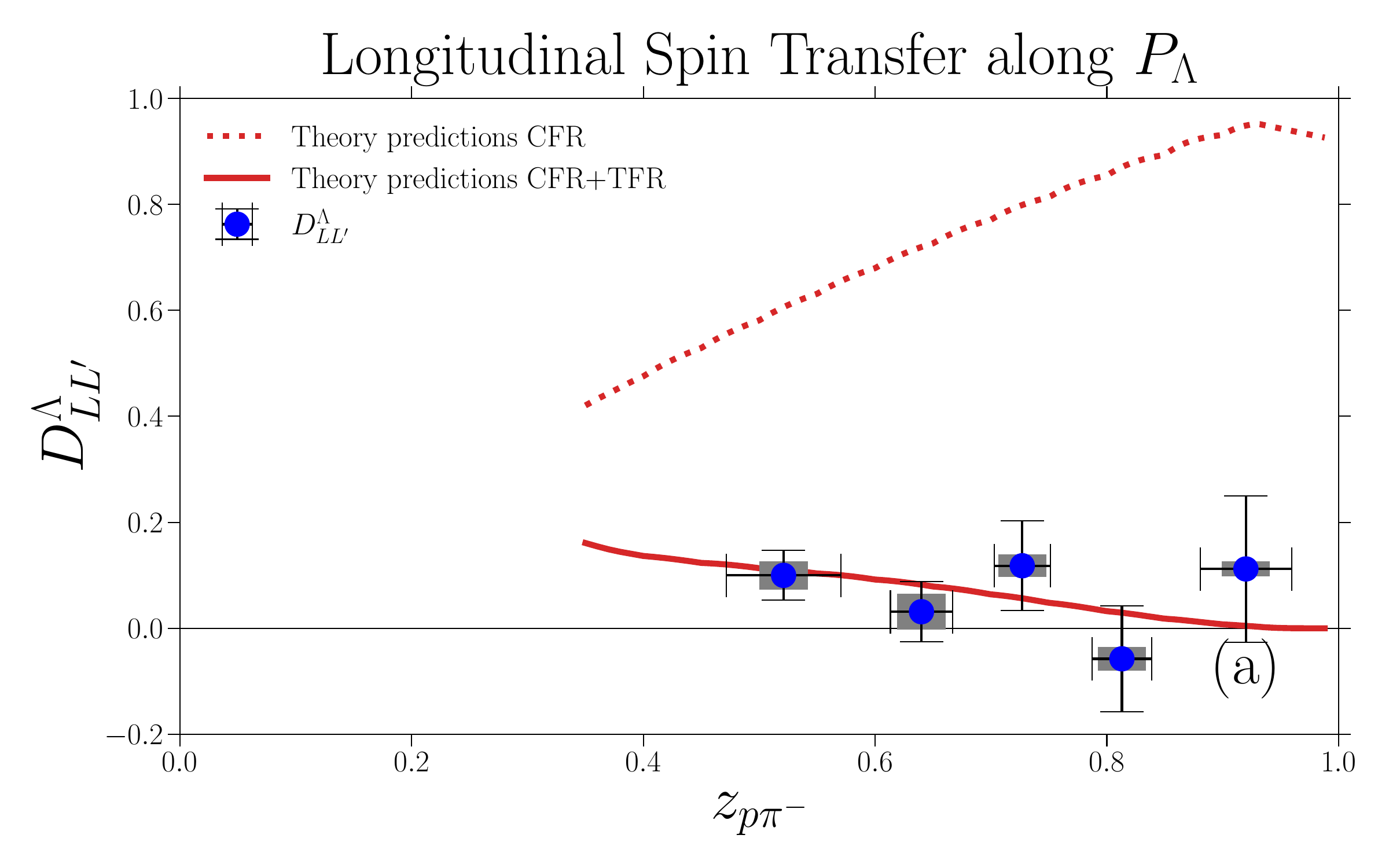}
\includegraphics[width=0.48\textwidth, left]{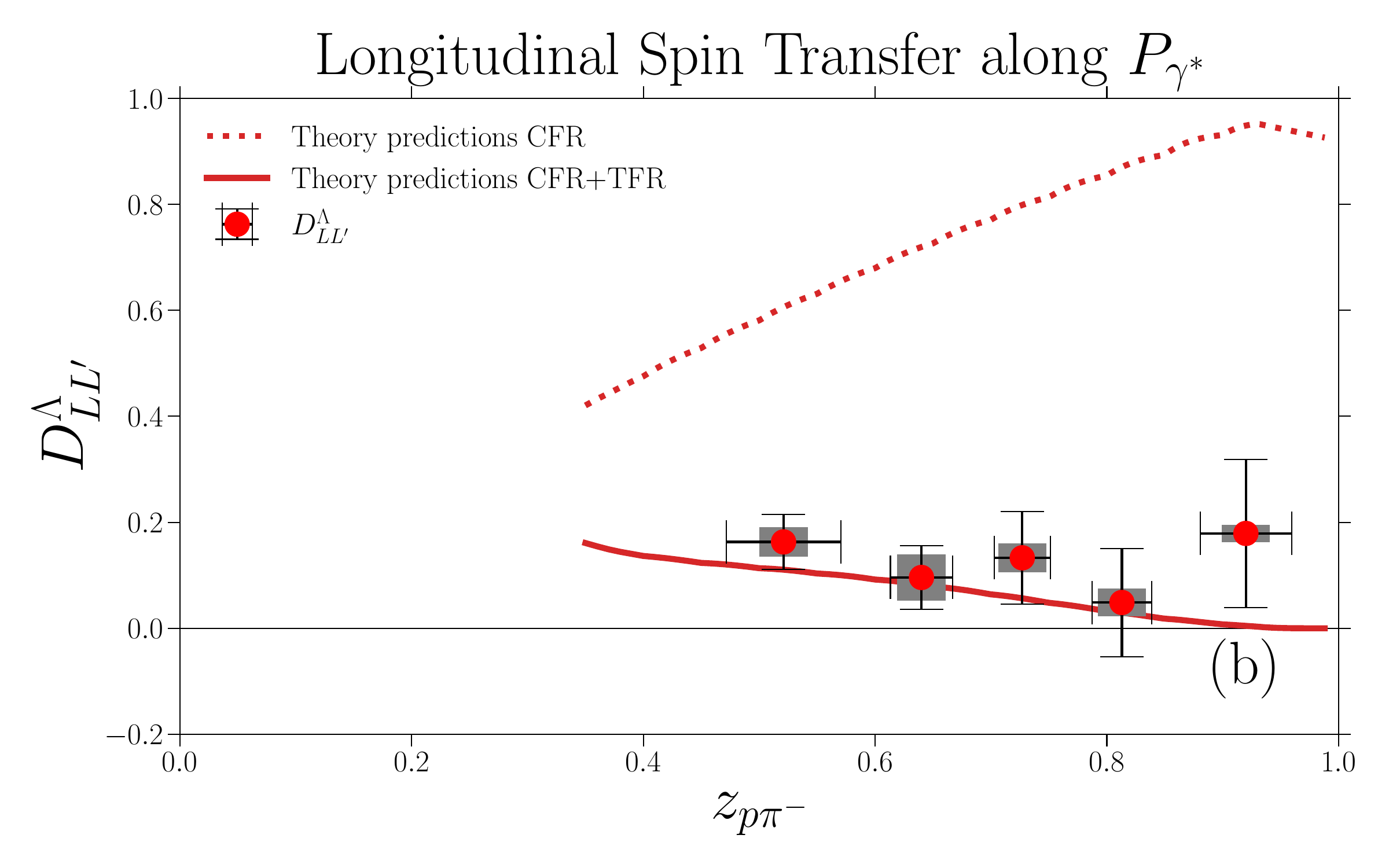}
\caption{Comparison of the extracted $D^{\Lambda}_{LL'}$ binned in $z_{p\pi^{-}}$ with predictions computed from FF parameterizations from pQCD~\cite{PhysRevD.57.5811} for the CFR and from a spectator quark-diquark model calculation for the TFR~\cite{zhao_prl_2025}.  Dashed curves represent CFR only and solid curves represent the sum of CFR and TFR contributions~\cite{zhaoprivate2025}.  The subpanels show results for $\Lambda$ polarization along the $\Lambda$ momentum (a) and along the virtual photon $\gamma^*$ momentum (b) in the $\gamma^*N$ CM frame.  Vertical uncertainties are statistical and horizontal uncertainties are given by the standard deviation within each bin.  Gray boxes show the systematic uncertainties.}
\label{fig:conclusions_theory_z_ppim}
\end{figure}

\begin{figure}[ht!]
\includegraphics[width=0.48\textwidth, left]{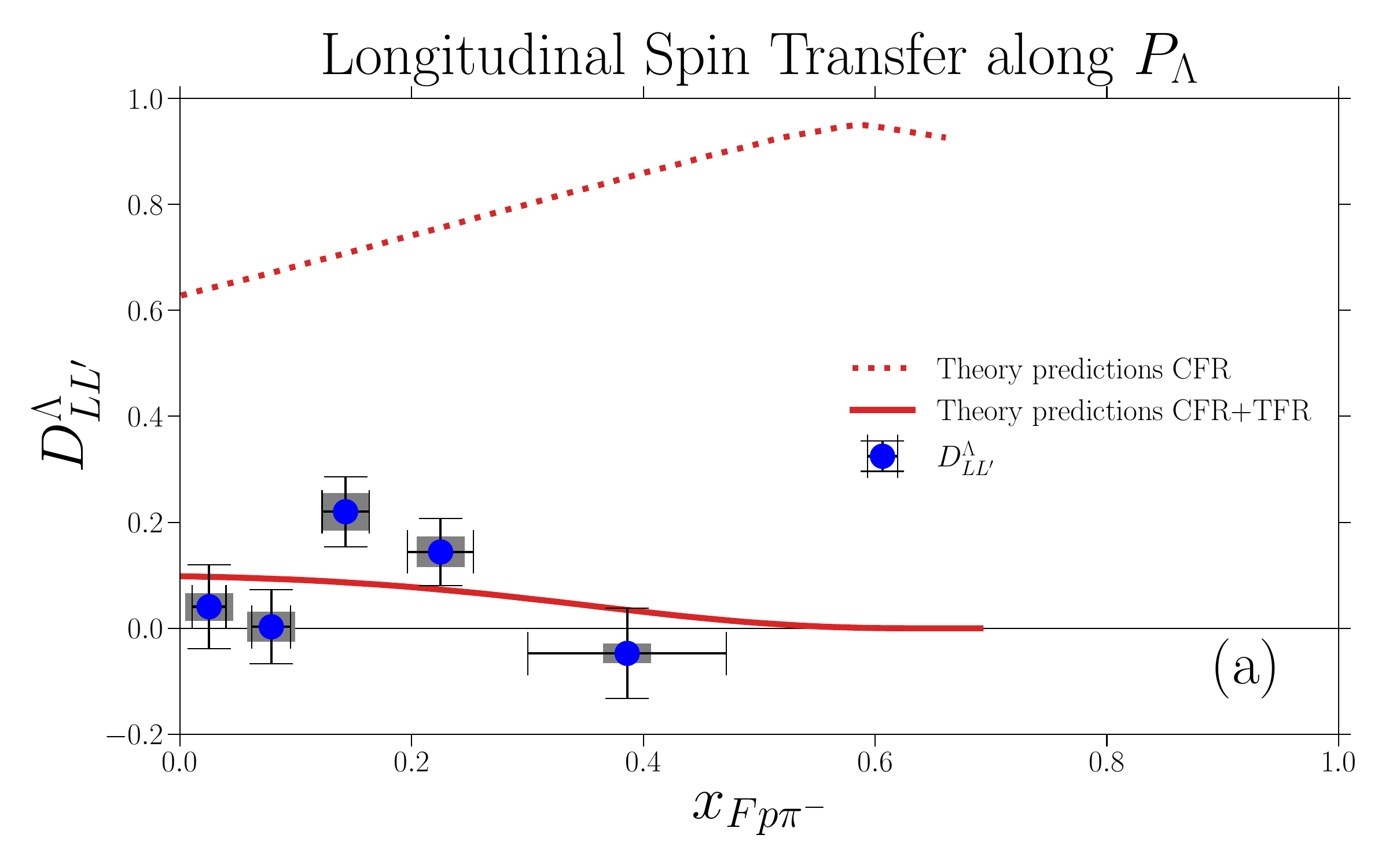}
\includegraphics[width=0.48\textwidth, left]{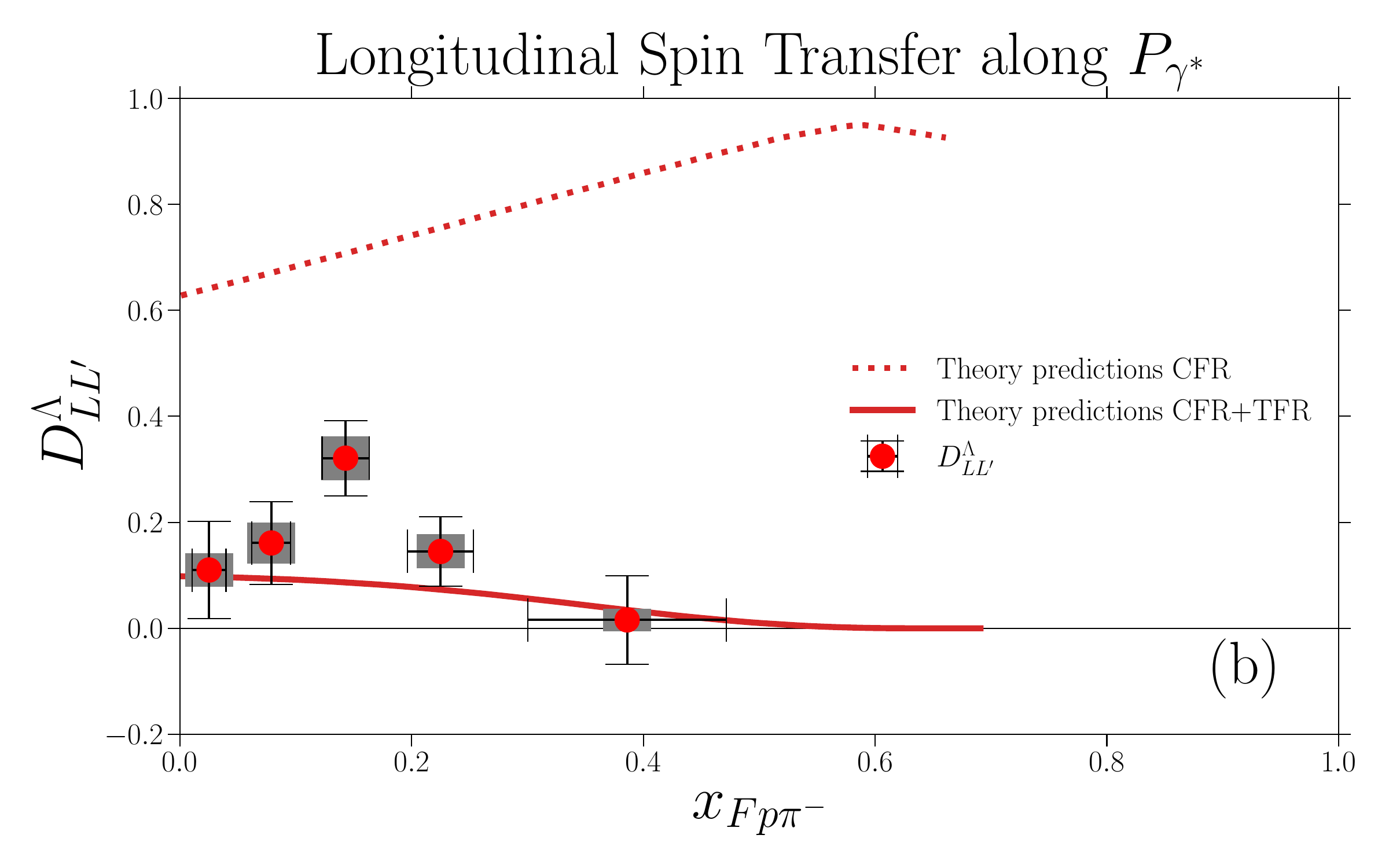}
\caption{Comparison of the extracted $D^{\Lambda}_{LL'}$ binned in $x_{F p\pi^{-}}$ with predictions computed from FF parameterizations from pQCD~\cite{PhysRevD.57.5811} for the CFR and from a spectator quark-diquark model calculation for the TFR~\cite{zhao_prl_2025}.  Dashed curves represent CFR only and solid curves represent the sum of CFR and TFR contributions~\cite{zhaoprivate2025}.  The subpanels show results for $\Lambda$ polarization along the $\Lambda$ momentum (a) and along the virtual photon $\gamma^*$ momentum (b) in the $\gamma^*N$ CM frame.  Vertical uncertainties are statistical and horizontal uncertainties are given by the standard deviation within each bin.  Gray boxes show the systematic uncertainties.}
\label{fig:conclusions_theory_xF_ppim}
\end{figure}

There is also a dilution effect from TFR $\Lambda$s that form from the remnant diquark system in the target nucleon.  $\Lambda$s originating from the TFR should not carry any spin transfer since the electron spin is only imparted to the struck diquark.  Hence, one expects TFR contributions to have zero spin transfer and to dilute any spin transfer from the CFR.  Since the $x_{F p\pi^{-}}>0$ cut does not guarantee that the sample is purely CFR $\Lambda$s, our results may be affected by some inevitable contamination from the TFR, and this is in fact observed by comparison with theory predictions~\cite{zhao_prl_2025} including both CFR and TFR contributions as shown in Figs.~\ref{fig:conclusions_theory_z_ppim} and~\ref{fig:conclusions_theory_xF_ppim}.  The CFR predictions were obtained with parameterizations of the $\Lambda$ FFs $D^{\Lambda}_{1,a}$ and $G^{\Lambda}_{1,a}$ inferred from a perturbative QCD (pQCD) fit to $e^{+}e^{-}$ annihilation data~\cite{PhysRevD.57.5811}, while a spectator quark-diquark model calculation was used to obtain the TFR predictions~\cite{zhao_prl_2025}.  Predictions were calculated for the mean kinematic values of our dataset at $Q^2=2.13$~GeV$^2$ and $x=0.25$.

%%% Fig: Lambda Parents %%%

%%% Fig: Comparison of results with theory predictions from Xiaoyan %%%

%%% Fig: Comparison of results with theory predictions from Xiaoyan %%%

The comparison of our observations with the combined CFR and TFR model predictions indicates firstly that our data contains a large TFR contribution even at high $x_{F p\pi^{-}}$ and hence the TFR and CFR contributions are not cleanly separable by a simple cut on $x_{F p\pi^{-}}$.  Furthermore, looking at the $x_{F p\pi^-}$ distribution, one may also notice a peak in the data around $x_{F p\pi^-}\simeq0.2$ which is not captured by the model.  This indicates that the possibility that there is more nuanced physics underlying our data than is described by the simple parameterizations of the polarized FF $G^{\Lambda}_{1,a}$ from measurements of the unpolarized FF $D^{\Lambda}_{1,a}$ employed in Ref.~\cite{PhysRevD.57.5811}.

This will require future measurements of SIDIS CFR and TFR quantities at these energy scales to carefully account for the relative contributions from each production mechanism.  Our measurement also constrains the extraction of related $\Lambda$ Fracture Functions (FrFs)~\cite{Trentadue1994FractureFA,Anselmino2011SIDISIT}, which describe the TFR production mechanism.  A FrF is typically expressed as a function of $x$ and $\zeta_h = \frac{P^-_h}{P^-}$, which is the negative light cone momentum component fraction for a hadron $h$ with 4-momentum $P_h$ produced off a nucleon target with 4-momentum $P$.  The reason $\zeta$ is introduced is that $z_h$ cannot distinguish between soft hadron emission and the target fragmentation region since it vanishes in the soft emission limit (hadron energy $E_h \rightarrow 0$) and in target fragmentation ($\theta_h=0$ where $\theta_h$ is the $\gamma^*N$ CM frame angle between $\vec{P}$ and $\vec{P}_{h}$), whereas $\zeta$ does not vanish in target fragmentation~\cite{Anselmino2011SIDISIT}.  Comparison with measurements of FFs from $e^{+}e^{-}$ annihilation, which obviously receives no TFR contribution, would allow one to deduce the FrFs.  Furthermore, because the spin transfer $D^{\Lambda}_{LL'}$ observable is sensitive to the production mechanism of the $\Lambda$, one may also infer loose constraints on the relative dominance of the CFR and TFR mechanisms in specific kinematic regions.  For example, due to helicity conservation the $\Lambda$ should carry the same helicity as the fragmenting quark as $z \rightarrow 1$, giving one a clear idea of the CFR contribution.  Conversely, in the valence region $x \rightarrow 1$ and $\zeta \rightarrow 1$, the TFR production mechanism highly favors unpolarized $\Lambda$ production due to the spin and flavor structure of the $\Lambda$ where the strange quark produced during fragmentation carries most of the $\Lambda$ spin.

\section*{Acknowledgements}
We thank Xiaoyan Zhao for the enlightening discussions and the model calculations.  We acknowledge the outstanding efforts of the staff of the Accelerator and the Physics Divisions at Jefferson Lab in making this experiment possible. This work was supported in part by the U.S. Department of Energy, the National Science Foundation (NSF), the Italian Istituto Nazionale di Fisica Nucleare (INFN), the French Centre National de la Recherche Scientifique (CNRS), the French Commissariat pour l'Energie Atomique, the UK Science and Technology Facilities Council, the National Research Foundation (NRF) of Korea, the HelmholtzForschungsakademie Hessen f\"{u}r FAIR (HFHF), the Ministry of Science and Higher Education of the Russian Federation, the Deutsche Forschungsgemeinschaft (DFG), the Science and Technology Facilities Council (STFC) grant ST/V001035/1, and the Chilean Agencia Nacional de Investigacion y Desarollo (ANID). The Southeastern Universities Research Association (SURA) operates the Thomas Jefferson National Accelerator Facility for the U.S. Department of Energy under Contract No. DE-AC05-06OR23177.  MM and AV are supported by the U.S. Department of Energy, Office of Science, Office of Nuclear Physics under Award No. DE-SC0019230.

% The \nocite command causes all entries in a bibliography to be printed out
% whether or not they are actually referenced in the text. This is appropriate
% for the sample file to show the different styles of references, but authors
% most likely will not want to use it.
% \nocite{*}

\FloatBarrier

\bibliographystyle{unsrt} % lists in order of appearance
\bibliography{refs}% Produces the bibliography via BibTeX.

\clearpage
\onecolumngrid
\appendix

\section{Tables} \label{app:Tables}
Here we present breakdowns of our systematic uncertainties according to their various sources in Tables~\ref{table:systematics_summary_ct1_z_ppim}-\ref{table:systematics_summary_ct2_xF_ppim}, our final results for $D^{\Lambda}_{LL'}$ in Tables~\ref{table:results_hb_ct1_z_ppim}-\ref{table:results_hb_ct2_xF_ppim}, and the kinematic means in Tables~\ref{table:results_kinematic_means_mass_ppim}-\ref{table:results_kinematic_means_xF_ppim} for each of our binning schemes and each choice of $\Lambda$ polarization axis.  The sources of systematic uncertainty are denoted as follows: $\Delta_{\alpha_{\Lambda}}$ refers to the systematic from the $\Lambda$ decay asymmetry parameter, $\Delta_{pol}$ refers to the systematic from the beam polarization, $\Delta_{Fit}$ refers to the systematic from the mass fit, $\Delta_{Inj}$ refers to the the systematic from the asymmetry injection study, and $\Delta_{\cos{\phi_{\Lambda}}}$ refers to the systematic from the additional injection study in different regions of $\phi_{\Lambda}$.  Uncertainties on the kinematic variables are taken as the standard deviation of the values in a given bin.
\FloatBarrier

%---------- SYSTEMATICS TABLES ----------%
%%% TABLE: SYSTEMATICS SUMMARY z_ppim CT1
\begin{table*}[] % ht!]
\centering
\begin{filecontents*}{tables/systematics/breakdown/aggregate___binvar_z_ppim__fitvar_costheta1__method_HB__z_ppim_0.0_1.0__systematics.pdf.csv}
bin,x,xerr,alphalambda,beam,inj,fit,cosphi
1,0.521,0.050,0.00120,0.00360,0.0083,0.0010,0.0253
2,0.640,0.027,0.00038,0.00113,0.0036,0.0132,0.0308
3,0.727,0.024,0.00141,0.00424,0.0072,0.0085,0.0175
4,0.813,0.026,0.00069,0.00207,0.0090,0.0019,0.0203
5,0.920,0.040,0.00135,0.00405,0.0044,0.0000,0.0128
\end{filecontents*}
\csvreader[
        head to column names,
        before reading = \begin{center}\sisetup{table-number-alignment=center},
        tabular = c c c c c c c c,
        table head = \toprule \textbf{Bin} & \textbf{$\braket{z_{p\pi^{-}}}$} & \textbf{$\Delta_{\alpha_{\Lambda}}$} & \textbf{$\Delta_{pol}$} & \textbf{$\Delta_{Fit}$} & \textbf{$\Delta_{Inj}$} & \textbf{$\Delta_{\cos{\phi_{\Lambda}}}$} \\\midrule,
        table foot = \bottomrule,
        after reading = \end{center},
]{tables/systematics/breakdown/aggregate___binvar_z_ppim__fitvar_costheta1__method_HB__z_ppim_0.0_1.0__systematics.pdf.csv}{}{$\bin$ & $\x\pm\xerr$ & $\alphalambda$ & $\beam$ & $\fit$ & $\inj$ & $\cosphi$}
\caption{Systematic uncertainties on $D_{LL'}^{\Lambda}$ for $\cos{\theta_{pL'}}$ along $P_{\Lambda}$ binned in $z_{p\pi^{-}}$.}
\label{table:systematics_summary_ct1_z_ppim}
\end{table*}

%%% TABLE: SYSTEMATICS SUMMARY z_ppim CT2
\begin{table*}[] % ht!]
\centering
\begin{filecontents*}{tables/systematics/breakdown/aggregate___binvar_z_ppim__fitvar_costheta2__method_HB__z_ppim_0.0_1.0__systematics.pdf.csv}
bin,x,xerr,alphalambda,beam,inj,fit,cosphi
1,0.521,0.050,0.00196,0.00587,0.0116,0.0013,0.0241
2,0.640,0.027,0.00115,0.00345,0.0087,0.0276,0.0329
3,0.727,0.024,0.00160,0.00479,0.0107,0.0113,0.0214
4,0.813,0.026,0.00058,0.00175,0.0166,0.0071,0.0189
5,0.920,0.040,0.00215,0.00645,0.0066,0.0000,0.0133
\end{filecontents*}
\csvreader[
        head to column names,
        before reading = \begin{center}\sisetup{table-number-alignment=center},
        tabular = c c c c c c c c,
        table head = \toprule \textbf{Bin} & \textbf{$\braket{z_{p\pi^{-}}}$} & \textbf{$\Delta_{\alpha_{\Lambda}}$} & \textbf{$\Delta_{pol}$} & \textbf{$\Delta_{Fit}$} & \textbf{$\Delta_{Inj}$} & \textbf{$\Delta_{\cos{\phi_{\Lambda}}}$} \\\midrule,
        table foot = \bottomrule,
        after reading = \end{center},
]{tables/systematics/breakdown/aggregate___binvar_z_ppim__fitvar_costheta2__method_HB__z_ppim_0.0_1.0__systematics.pdf.csv}{}{$\bin$ & $\x\pm\xerr$ & $\alphalambda$ & $\beam$ & $\fit$ & $\inj$ & $\cosphi$}
\caption{Systematic uncertainties on $D_{LL'}^{\Lambda}$ for $\cos{\theta_{pL'}}$ along $P_{\gamma^{*}}$ binned in $z_{p\pi^{-}}$.}
\label{table:systematics_summary_ct2_z_ppim}
\end{table*}

%%% TABLE: SYSTEMATICS SUMMARY xF_ppim CT1
\begin{table*}[] % ht!]
\centering
\begin{filecontents*}{tables/systematics/breakdown/aggregate___binvar_xF_ppim__fitvar_costheta1__method_HB__xF_ppim_0.0_1.0__systematics.pdf.csv}
bin,x,xerr,alphalambda,beam,inj,fit,cosphi
1,0.025,0.015,0.00049,0.00146,0.0056,0.0029,0.0254
2,0.079,0.017,0.00003,0.00010,0.0085,0.0041,0.0270
3,0.143,0.020,0.00264,0.00793,0.0041,0.0192,0.0284
4,0.225,0.029,0.00173,0.00519,0.0105,0.0188,0.0185
5,0.386,0.086,0.00057,0.00170,0.0108,0.0066,0.0135
\end{filecontents*}
\csvreader[
        head to column names,
        before reading = \begin{center}\sisetup{table-number-alignment=center},
        tabular = c c c c c c c c,
        table head = \toprule \textbf{Bin} & \textbf{$\braket{x_{F p\pi^{-}}}$} & \textbf{$\Delta_{\alpha_{\Lambda}}$} & \textbf{$\Delta_{pol}$} & \textbf{$\Delta_{Fit}$} & \textbf{$\Delta_{Inj}$} & \textbf{$\Delta_{\cos{\phi_{\Lambda}}}$} \\\midrule,
        table foot = \bottomrule,
        after reading = \end{center},
]{tables/systematics/breakdown/aggregate___binvar_xF_ppim__fitvar_costheta1__method_HB__xF_ppim_0.0_1.0__systematics.pdf.csv}{}{$\bin$ & $\x\pm\xerr$ & $\alphalambda$ & $\beam$ & $\fit$ & $\inj$ & $\cosphi$}
\caption{Systematic uncertainties on $D_{LL'}^{\Lambda}$ for $\cos{\theta_{pL'}}$ along $P_{\Lambda}$ binned in $x_{F p\pi^{-}}$.}
\label{table:systematics_summary_ct1_xF_ppim}
\end{table*}

%%% TABLE: SYSTEMATICS SUMMARY xF_ppim CT2
\begin{table*}[] % ht!]
\centering
\begin{filecontents*}{tables/systematics/breakdown/aggregate___binvar_xF_ppim__fitvar_costheta2__method_HB__xF_ppim_0.0_1.0__systematics.pdf.csv}
bin,x,xerr,alphalambda,beam,inj,fit,cosphi
1,0.025,0.015,0.00132,0.00395,0.0082,0.0070,0.0292
2,0.079,0.017,0.00193,0.00580,0.0129,0.0187,0.0310
3,0.143,0.020,0.00386,0.01160,0.0045,0.0257,0.0301
4,0.225,0.029,0.00174,0.00522,0.0150,0.0189,0.0206
5,0.386,0.086,0.00019,0.00057,0.0128,0.0093,0.0143
\end{filecontents*}
\csvreader[
        head to column names,
        before reading = \begin{center}\sisetup{table-number-alignment=center},
        tabular = c c c c c c c c,
        table head = \toprule \textbf{Bin} & \textbf{$\braket{x_{F p\pi^{-}}}$} & \textbf{$\Delta_{\alpha_{\Lambda}}$} & \textbf{$\Delta_{pol}$} & \textbf{$\Delta_{Fit}$} & \textbf{$\Delta_{Inj}$} & \textbf{$\Delta_{\cos{\phi_{\Lambda}}}$} \\\midrule,
        table foot = \bottomrule,
        after reading = \end{center},
]{tables/systematics/breakdown/aggregate___binvar_xF_ppim__fitvar_costheta2__method_HB__xF_ppim_0.0_1.0__systematics.pdf.csv}{}{$\bin$ & $\x\pm\xerr$ & $\alphalambda$ & $\beam$ & $\fit$ & $\inj$ & $\cosphi$}
\caption{Systematic uncertainties on $D_{LL'}^{\Lambda}$ for $\cos{\theta_{pL'}}$ along $P_{\gamma^{*}}$ binned in $x_{F p\pi^{-}}$.}
\label{table:systematics_summary_ct2_xF_ppim}
\end{table*}

%----------------------------------------%

%--------------- Results ----------------%
%%% z_ppim Binning %%%
\begin{table*}[t]
\centering
\begin{filecontents*}{tables/results/aggregate___binvar_z_ppim__fitvar_costheta1__method_HB__z_ppim_0.0_1.0.pdf.csv}
bin,x,y,xerr,yerr,xerrsyst,yerrsyst
1,0.521,0.100,0.050,0.047,0,0.027
2,0.640,0.031,0.027,0.057,0,0.034
3,0.727,0.118,0.024,0.085,0,0.021
4,0.813,0.000,0.026,0.100,0,0.022
5,0.920,0.112,0.040,0.138,0,0.014
\end{filecontents*}
\csvreader[
        head to column names,
        before reading = \begin{center}\sisetup{table-number-alignment=center},
        tabular = c cc,
        table head = \toprule \textbf{Bin} & \textbf{$\braket{z_{p\pi^{-}}}$} & \textbf{Extracted Asymmetry} \\\midrule,
        table foot = \bottomrule,
        after reading = \end{center},
]{tables/results/aggregate___binvar_z_ppim__fitvar_costheta1__method_HB__z_ppim_0.0_1.0.pdf.csv}
{}{\bin & \x\hspace{0.1cm}$\pm$\hspace{0.1cm}\xerr & \y\hspace{0.1cm}$\pm$\hspace{0.1cm}\yerr\hspace{0.1cm}\text{(stat)}\hspace{0.1cm}$\pm$\hspace{0.1cm}\yerrsyst\hspace{0.1cm}\text{(syst)}}
\caption{$D_{LL'}^{\Lambda}$ binned in $z_{p\pi^{-}}$ for $\cos{\theta_{pL'}}$ along $\Vec{P}_{\Lambda}$.}
\label{table:results_hb_ct1_z_ppim}
\end{table*}

\begin{table*}[t]
\centering
\begin{filecontents*}{tables/results/aggregate___binvar_z_ppim__fitvar_costheta2__method_HB__z_ppim_0.0_1.0.pdf.csv}
bin,x,y,xerr,yerr,xerrsyst,yerrsyst
1,0.521,0.163,0.050,0.052,0,0.028
2,0.640,0.096,0.027,0.060,0,0.044
3,0.727,0.133,0.024,0.087,0,0.027
4,0.813,0.049,0.026,0.102,0,0.026
5,0.920,0.179,0.040,0.140,0,0.016
\end{filecontents*}
\csvreader[
        head to column names,
        before reading = \begin{center}\sisetup{table-number-alignment=center},
        tabular = c cc,
        table head = \toprule \textbf{Bin} & \textbf{$\braket{z_{p\pi^{-}}}$} & \textbf{Extracted Asymmetry} \\\midrule,
        table foot = \bottomrule,
        after reading = \end{center},
]{tables/results/aggregate___binvar_z_ppim__fitvar_costheta2__method_HB__z_ppim_0.0_1.0.pdf.csv}
{}{\bin & \x\hspace{0.1cm}$\pm$\hspace{0.1cm}\xerr & \y\hspace{0.1cm}$\pm$\hspace{0.1cm}\yerr\hspace{0.1cm}\text{(stat)}\hspace{0.1cm}$\pm$\hspace{0.1cm}\yerrsyst\hspace{0.1cm}\text{(syst)}}
\caption{$D_{LL'}^{\Lambda}$ binned in $z_{p\pi^{-}}$ for $\Vec{P}_{\Lambda}$ along $\Vec{P}_{\gamma^*}$.}
\label{table:results_hb_ct2_z_ppim}
\end{table*}

%%% xF_ppim Binning %%%
\begin{table*}[t]
\centering
\begin{filecontents*}{tables/results/aggregate___binvar_xF_ppim__fitvar_costheta1__method_HB__xF_ppim_0.0_1.0.pdf.csv}
bin,x,y,xerr,yerr,xerrsyst,yerrsyst
1,0.025,0.041,0.015,0.079,0,0.026
2,0.079,0.003,0.017,0.070,0,0.029
3,0.143,0.220,0.020,0.066,0,0.036
4,0.225,0.144,0.029,0.063,0,0.029
5,0.386,0.000,0.086,0.085,0,0.019
\end{filecontents*}
\csvreader[
        head to column names,
        before reading = \begin{center}\sisetup{table-number-alignment=center},
        tabular = c cc,
        table head = \toprule \textbf{Bin} & \textbf{$\braket{x_{F p\pi^{-}}}$} & \textbf{Extracted Asymmetry} \\\midrule,
        table foot = \bottomrule,
        after reading = \end{center},
]{tables/results/aggregate___binvar_xF_ppim__fitvar_costheta1__method_HB__xF_ppim_0.0_1.0.pdf.csv}
{}{\bin & \x\hspace{0.1cm}$\pm$\hspace{0.1cm}\xerr & \y\hspace{0.1cm}$\pm$\hspace{0.1cm}\yerr\hspace{0.1cm}\text{(stat)}\hspace{0.1cm}$\pm$\hspace{0.1cm}\yerrsyst\hspace{0.1cm}\text{(syst)}}
\caption{$D_{LL'}^{\Lambda}$ binned in $x_{F p\pi^{-}}$ for $\cos{\theta_{pL'}}$ along $\Vec{P}_{\Lambda}$.}
\label{table:results_hb_ct1_xF_ppim}
\end{table*}

\begin{table*}[t]
\centering
\begin{filecontents*}{tables/results/aggregate___binvar_xF_ppim__fitvar_costheta2__method_HB__xF_ppim_0.0_1.0.pdf.csv}
bin,x,y,xerr,yerr,xerrsyst,yerrsyst
1,0.025,0.110,0.015,0.091,0,0.031
2,0.079,0.161,0.017,0.078,0,0.039
3,0.143,0.321,0.020,0.071,0,0.042
4,0.225,0.145,0.029,0.065,0,0.032
5,0.386,0.016,0.086,0.083,0,0.021
\end{filecontents*}
\csvreader[
        head to column names,
        before reading = \begin{center}\sisetup{table-number-alignment=center},
        tabular = c cc,
        table head = \toprule \textbf{Bin} & \textbf{$\braket{x_{F p\pi^{-}}}$} & \textbf{Extracted Asymmetry} \\\midrule,
        table foot = \bottomrule,
        after reading = \end{center},
]{tables/results/aggregate___binvar_xF_ppim__fitvar_costheta2__method_HB__xF_ppim_0.0_1.0.pdf.csv}
{}{\bin & \x\hspace{0.1cm}$\pm$\hspace{0.1cm}\xerr & \y\hspace{0.1cm}$\pm$\hspace{0.1cm}\yerr\hspace{0.1cm}\text{(stat)}\hspace{0.1cm}$\pm$\hspace{0.1cm}\yerrsyst\hspace{0.1cm}\text{(syst)}}
\caption{$D_{LL'}^{\Lambda}$ binned in $x_{F p\pi^{-}}$ for $\cos{\theta_{pL'}}$ along $\Vec{P}_{\gamma^*}$.}
\label{table:results_hb_ct2_xF_ppim}
\end{table*}

%----------------------------------------%

%---------- Kinematics Tables ----------%

%%% Overall Kinematic Means %%%
\begin{table*}[b] % ht!]
\centering
\begin{filecontents*}{tables/results/kinematic_means/kinematics_mass_ppim.csv}
bin,massppim,massppimerr,Q,Qerr,W,Werr,y,yerr,x,xerr,xFppim,xFppimerr,zppim,zppimerr
1,1.12,0.01,2.21,1.18,2.79,0.47,0.469,0.139,0.247,0.117,0.199,0.140,0.705,0.143
\end{filecontents*}
\csvreader[
        head to column names,
        before reading = \begin{center}\sisetup{table-number-alignment=center},
        tabular = c c c c c c c c,
        table head = \toprule \textbf{Bin} & \textbf{$\braket{M_{p\pi^{-}}}$}~(GeV) & \textbf{$\braket{Q^{2}}$}~(GeV$^2$) & \textbf{$\braket{W}$}~(GeV) & \textbf{$\braket{y}$} & \textbf{$\braket{x}$} & \textbf{$\braket{z_{p\pi^{-}}}$} & \textbf{$\braket{x_{F p\pi^{-}}}$} \\\midrule,
        table foot = \bottomrule,
        after reading = \end{center},
]{tables/results/kinematic_means/kinematics_mass_ppim.csv}
{}{\bin & \massppim$\pm$\massppimerr & \Q$\pm$\Qerr & \W$\pm$\Werr & \y$\pm$\yerr & \x$\pm$\xerr & \zppim$\pm$\zppimerr & \xFppim$\pm$\xFppimerr}
% \bin & \massppim$\pm$\massppimerr & \Q$\pm$\Qerr & \W$\pm$\Werr & \y$\pm$\yerr & \x$\pm$\xerr & \xFppim$\pm$\xFppimerr & \zppim$\pm$\zppimerr
\caption{Kinematic means in the $\Lambda$ signal region averaged over the entire dataset.}
\label{table:results_kinematic_means_mass_ppim}
\end{table*}

%%% z_ppim Kinematic Means %%%
\begin{table*}[t] % ht!]
\centering
\begin{filecontents*}{tables/results/kinematic_means/kinematics_z_ppim.csv}
bin,massppim,massppimerr,Q,Qerr,W,Werr,y,yerr,x,xerr,xFppim,xFppimerr,zppim,zppimerr
1,1.12,0.01,2.17,1.14,3.27,0.33,0.607,0.103,0.182,0.091,0.091,0.066,0.521,0.050
2,1.12,0.01,2.32,1.28,2.89,0.35,0.499,0.105,0.236,0.113,0.158,0.095,0.640,0.027
3,1.12,0.01,2.31,1.25,2.68,0.36,0.439,0.106,0.267,0.120,0.208,0.117,0.727,0.024
4,1.12,0.01,2.21,1.15,2.51,0.34,0.390,0.101,0.287,0.117,0.260,0.133,0.813,0.026
5,1.12,0.01,2.02,1.03,2.37,0.30,0.345,0.091,0.294,0.107,0.339,0.147,0.920,0.040
\end{filecontents*}
\csvreader[
        head to column names,
        before reading = \begin{center}\sisetup{table-number-alignment=center},
        tabular = c c c c c c c c,
        table head = \toprule \textbf{Bin} & \textbf{$\braket{M_{p\pi^{-}}}$}~(GeV) & \textbf{$\braket{Q^{2}}$}~(GeV$^2$) & \textbf{$\braket{W}$}~(GeV) & \textbf{$\braket{y}$} & \textbf{$\braket{x}$} & \textbf{$\braket{z_{p\pi^{-}}}$} & \textbf{$\braket{x_{F p\pi^{-}}}$} \\\midrule,
        table foot = \bottomrule,
        after reading = \end{center},
]{tables/results/kinematic_means/kinematics_z_ppim.csv}
{}{\bin & \massppim$\pm$\massppimerr & \Q$\pm$\Qerr & \W$\pm$\Werr & \y$\pm$\yerr & \x$\pm$\xerr & \zppim$\pm$\zppimerr & \xFppim$\pm$\xFppimerr}
% \bin & \massppim$\pm$\massppimerr & \Q$\pm$\Qerr & \W$\pm$\Werr & \y$\pm$\yerr & \x$\pm$\xerr & \xFppim$\pm$\xFppimerr & \zppim$\pm$\zppimerr
\caption{Kinematic means in the $\Lambda$ signal region for the $z_{p\pi^{-}}$ binning.}
\label{table:results_kinematic_means_z_ppim}
\end{table*}

%%% xF_ppim Kinematic Means %%%
\begin{table*}[t] % ht!]
\centering
\begin{filecontents*}{tables/results/kinematic_means/kinematics_xF_ppim.csv}
bin,massppim,massppimerr,Q,Qerr,W,Werr,y,yerr,x,xerr,xFppim,xFppimerr,zppim,zppimerr
1,1.12,0.01,2.27,1.21,2.87,0.49,0.495,0.143,0.242,0.121,0.025,0.015,0.588,0.123
2,1.12,0.01,2.24,1.20,2.83,0.48,0.482,0.142,0.245,0.119,0.079,0.017,0.625,0.122
3,1.12,0.01,2.20,1.17,2.78,0.48,0.468,0.141,0.248,0.117,0.143,0.020,0.669,0.120
4,1.12,0.01,2.20,1.18,2.75,0.47,0.456,0.138,0.252,0.117,0.225,0.029,0.726,0.114
5,1.12,0.01,2.18,1.16,2.76,0.44,0.458,0.130,0.247,0.114,0.386,0.086,0.825,0.101
\end{filecontents*}
\csvreader[
        head to column names,
        before reading = \begin{center}\sisetup{table-number-alignment=center},
        tabular = c c c c c c c c,
        table head = \toprule \textbf{Bin} & \textbf{$\braket{M_{p\pi^{-}}}$}~(GeV) & \textbf{$\braket{Q^{2}}$}~(GeV$^2$) & \textbf{$\braket{W}$}~(GeV) & \textbf{$\braket{y}$} & \textbf{$\braket{x}$} & \textbf{$\braket{z_{p\pi^{-}}}$} & \textbf{$\braket{x_{F p\pi^{-}}}$} \\\midrule,
        table foot = \bottomrule,
        after reading = \end{center},
]{tables/results/kinematic_means/kinematics_xF_ppim.csv}
{}{\bin & \massppim$\pm$\massppimerr & \Q$\pm$\Qerr & \W$\pm$\Werr & \y$\pm$\yerr & \x$\pm$\xerr & \zppim$\pm$\zppimerr & \xFppim$\pm$\xFppimerr}
% \bin & \massppim$\pm$\massppimerr & \Q$\pm$\Qerr & \W$\pm$\Werr & \y$\pm$\yerr & \x$\pm$\xerr & \xFppim$\pm$\xFppimerr & \zppim$\pm$\zppimerr
\caption{Kinematic means in the $\Lambda$ signal region for the $x_{F p\pi^{-}}$ binning.}
\label{table:results_kinematic_means_xF_ppim}
\end{table*}

%----------------------------------------%

\FloatBarrier

\end{document}